\documentclass[twocolumn,nofootinbib,showpacs,groupedaddress,
               superscriptaddress]{revtex4}
\usepackage{amssymb} 
\usepackage{graphicx}
\usepackage{amsmath}

\def\e{{\rm e}}

\newcommand{\be}{\begin{equation}}
\newcommand{\ee}{\end{equation}}
\newcommand{\bea}{\begin{eqnarray}}
\newcommand{\eea}{\end{eqnarray}}
\newcommand{\bg}{\begin{gather}}
\newcommand{\eg}{\end{gather}}
\newcommand{\bseq}{\begin{subequations}}
\newcommand{\eseq}{\end{subequations}}

\renewcommand{\ln}{\mathop{\rm ln}\nolimits}

\renewcommand{\Im}{\mathop{\rm Im}\nolimits}
\renewcommand{\Re}{\mathop{\rm Re}\nolimits}

\newcommand{\bra}[1]{\langle #1 |}
\newcommand{\ket}[1]{| #1 \rangle}
\newcommand{\braket}[2]{\langle #1 |#2 \rangle}
\newcommand{\bpm}{\begin{pmatrix}}
\newcommand{\epm}{\end{pmatrix}}
\makeatletter

\def\lsim{\compoundrel<\over\sim}
\def\compoundrel#1\over#2{\mathpalette\compoundreL{{#1}\over{#2}}}
\def\compoundreL#1#2{\compoundREL#1#2}
\def\compoundREL#1#2\over#3{\mathrel
         {\vcenter{\hbox{$\m@th\buildrel{#1#2}\over{#1#3}$}}}}
\makeatother

\begin{document}

\title{Complex trajectories in chaotic dynamical tunneling} 
\author{D.G.~Levkov}\email{levkov@ms2.inr.ac.ru}
\affiliation{Institute for Nuclear Research of the Russian Academy of
  Sciences, 60th October Anniversary prospect 7a, Moscow 117312,
  Russia} 

\author{A.G.~Panin}\email{panin@ms2.inr.ac.ru}
\affiliation{Institute for Nuclear Research of the Russian Academy of
  Sciences, 60th October Anniversary prospect 7a, Moscow 117312,
  Russia} 
\affiliation{Moscow Institute of Physics and Technology, Institutskii
  per. 9, Dolgoprudny 141700, Moscow Region, Russia}

\author{S.M.~Sibiryakov}\email{Sergey.Sibiryakov@cern.ch}
\affiliation{Institute for Nuclear Research of the Russian Academy of
  Sciences, 60th October Anniversary prospect 7a, Moscow 117312,
  Russia} 
\affiliation{Theory Group, Physics Department, CERN, CH-1211 Geneva
  23, Switzerland}

\date{\today}

\begin{abstract}
We develop the semiclassical
method of complex trajectories in application
to chaotic dynamical tunneling.
First, we suggest a
systematic numerical technique for obtaining complex tunneling
trajectories by the gradual deformation of the classical
ones. This provides a natural classification of the tunneling
solutions. Second, we present a heuristic procedure for sorting out the
least suppressed trajectory.
As an illustration, we apply our technique to the process of chaotic
tunneling in a quantum
mechanical model with two degrees of freedom.
Our analysis reveals rich
dynamics of the system. At the classical level, there exists an
infinite set of unstable solutions forming a fractal structure. This
structure is inherited by the complex tunneling paths and plays the
central role in the semiclassical study.
The process we consider exhibits the phenomenon of optimal tunneling:
the suppression exponent of
the tunneling probability has a
local minimum at a certain energy which is thus (locally) the optimal
energy for tunneling.
We test the proposed method by comparison of the semiclassical results
with the results of the exact quantum computations and find a good
agreement.
\end{abstract}



\maketitle

\section{Introduction}
\label{sec:1}
An intrinsic feature of quantum physics is the existence of processes 
which are forbidden  at the classical level. The text--book examples
of such processes are tunneling and over--barrier reflection in
one--dimensional quantum mechanics; more involved topics include
atom ionization processes~\cite{Perelomov}, chemical
reactions~\cite{Miller}, false vacuum decay in scalar field
theory~\cite{Coleman:1977py}, etc.
Generically, one introduces a certain parameter
$g^2$ (the Planck constant $\hbar$ in quantum mechanics, coupling
constant in field theory, etc.) measuring the magnitude of quantum
fluctuations, and finds that the probabilities ${\cal P}$ of classically
forbidden processes behave exponentially as $g^2\to 0$,
\be
\label{probab}
{\cal P}\simeq g^\gamma A\e^{-F/g^2}\;.
\ee
Here $F>0$ is the suppression exponent and the
dependence of the pre-exponential factor 
on $g$ is indicated explicitly. 
In this paper we adopt the term
 ``tunneling'' for
any process forbidden at the classical level.
This
includes, in particular, the cases of dynamical
tunneling~\cite{Miller,Heller:1981}, when the
exponential suppression of the process is not related to the existence
of a potential barrier.

A powerful tool for the study of tunneling
at small $g^2$ is provided by the semiclassical methods. Exploiting the
semiclassical approach, one reduces the problem of 
computing the tunneling probability to the problem of finding the
relevant 
solution to the classical equations of motion in the complex
domain~\cite{Perelomov,Miller,Wilkinson:Takada}, where both the time 
variable $t$ and dynamical coordinates are taken to be
complex. The
suppression exponent $F$ is then related to the 
classical action $S$  calculated along an appropriate contour
in the complex time plane. The shape of 
this contour, as well as the boundary conditions imposed on the
solution at $t\to \pm \infty$, are dictated by the quantum
numbers of the initial and final states. In the
simplest cases of under--barrier motion (one-dimensional
tunneling, false vacuum decay) the 
contour runs along the imaginary time  axis\footnote{In that case one
  usually introduces the real variable  $\tau=it$, which is called 
  Euclidean time.}, and the relevant solution is real along this axis.
Such a trajectory may be identified with the
``most probable escape path'' in the configuration
space~\cite{Banks:1973ps,Coleman:1977py}, which 
gives some understanding of the classically forbidden dynamics.  
In other cases, however, the passage of the system through the
classically forbidden region of the phase space cannot be treated
separately from the  
preceding and following real--time evolution, and the analysis of complex
tunneling solutions in the complex time domain is needed. This
happens, e.g., in the study of chemical reactions with definite
initial quantum state of reactants~\cite{Miller} or in the
investigation of the induced tunneling processes in field
theory~\cite{Rubakov:1992ec,Kuznetsov:1997az,Bezrukov:2003er,Levkov:2004tf}.
The semiclassical techniques based on genuinely complex
classical solutions received the common name of the method of complex
trajectories. 

It should be pointed out that the application of the above method 
might be highly
non--trivial. Major complications are related to the issues of
existence and uniqueness of tunneling trajectories, which are
basically the complex solutions to a certain boundary value
problem. First, it may occur that the 
boundary value problem at hand does not have any solutions at all (see, e.g.,
Ref.~\cite{Affleck:1980mp}). Second, there may exist many (sometimes an
infinite number of) solutions. Some of them may well be
unphysical and should be rejected. The identification of physical
solutions relies very much on the particular
properties of the system under consideration; presently there are no
trustworthy criteria applicable in general (see
Refs.~\cite{Berry:1972,Adachi:1986,Shudo:1996,Ribeiro:2004,Parisio:2005}
for the attempts to find such 
criteria). Even after the unphysical solutions are eliminated, the problem
remains to identify the solution(s) which yield the minimal suppression
exponent and therefore correspond to the dominant contribution to
the tunneling amplitude. 

The above difficulties become particularly pronounced in the 
case of tunneling in chaotic systems. Appearance of
chaos is generic for non--linear systems with many degrees of
freedom; the topics related to tunneling in the presence of
chaos were addressed in Refs.~\cite{Adachi:1986,Bohigas1993,
Doron1995,Mouchet2001,Shudo1995_1,Shudo:1996,Shudo2002,Onishi2003,
Ribeiro:2004}. 
The semiclassical analysis of chaotic tunneling is hindered by the
existence of an infinite number of semiclassical solutions which form
a fractal set in the complex phase 
space~\cite{Adachi:1986,Shudo1995_1,Onishi2003}. 
The direct analysis of this set  with the purpose of identifying the
physically relevant solutions  becomes
an elaborative task in the systems with many degrees of 
freedom~\cite{Heller2002}. Presently, the  
semiclassical analysis of the chaotic tunneling processes is limited
to the special cases when the phase space of the system can be
explicitly visualized~\cite{Adachi:1986,Shudo1995_1,Onishi2003}, or when the
small sub--class of 
periodic tunneling orbits is considered~\cite{Ribeiro:2004}. Development of
generic methods of classifying the semiclassical solutions
is of great importance~\cite{Heller2002}.

In this paper we present new method of obtaining and classifying the 
tunneling trajectories, which is applicable, in particular, in the case
of many--dimensional ($D\geq 2$) chaotic tunneling. Namely, we
consider the 
processes which proceed classically at some values of
the initial-state quantum numbers, and become exponentially suppressed
at other values. The technique presented below enables one to obtain
complex trajectories describing tunneling by starting from the real
classical solutions and changing gradually the quantum numbers of the
initial state. Our procedure has two advantages. First, it is generic
and numerically implementable. Second, it provides a  natural 
classification of tunneling trajectories based on the
analysis of their classical progenitors. The latter classification
suggests a heuristic method for sorting out the least suppressed
tunneling path. 

As an illustration we apply the above method to the problem of
scattering  in quantum mechanical model with two degrees of freedom. 
The process we study is a particular exemple of over-barrier
reflection. 
We calculate the suppression exponent for the reflection probability. 
As the test of the method we compare the semiclassical
results with the ``exact'' suppression exponent. The latter is
extracted from the exact wave function obtained by solving numerically
the Schr\"{o}dinger equation. The results of the two calculations  are
in good agreement.  

The system under consideration has two distinctive features, which are
inherent in two wide (intersecting but not 
necessarily identical) classes of tunneling problems. We believe that 
our model is a generic representative of both of these classes.

The first feature is chaoticity. In our model chaos manifests itself
at the classical level as follows. 
Consider the set of initial data giving rise to the classical
reflected trajectories. We will see that this set falls into an
infinite number of disconnected domains. The boundaries of these
domains correspond to trajectories which do not escape into the final
asymptotic region as time goes on, but get trapped in the interaction
region. The 
latter trajectories are unstable: small deviations from them lead to
either reflected or transmitted solutions. Increasing the resolution
of initial data reveals that the set of initial
data corresponding to the trapped trajectories forms a fractal similar
to the Cantor set. This is the hallmark of the so--called irregular (or
chaotic) scattering \cite{Eckhardt1986}.

The chaoticity of the classical dynamics has profound consequences for
the tunneling process. We will show that complex trajectories relevant
for over--barrier reflection are all trapped in the interaction region
and thus are unstable. Moreover, they inherit the fractal structure of
the classical trapped solutions which means, in particular, that their
number is infinite. The complex trajectory which contributes most to
the tunneling amplitude is a descendant of a certain unstable
classical solution lying on the boundary of the above fractal set. 

The chaoticity of the process manifests itself in the exact
quantum computations as well. We find that the quantum probability of
tunneling, instead of being smooth function of energy, exhibits large
irregular oscillations. Similar dependence of the tunneling
amplitude on the parameters of chaotic systems was reported previously in
Refs.~\cite{Bohigas1993,Doron1995,Shudo1995_1}. At first glance, this
behavior contradicts to the semiclassical formula
(\ref{probab}). One observes, however, that the oscillation period 
scales like $g^2$ when 
$g^2\to 0$, so that the oscillations become indiscernible in
the semiclassical limit. In order to extract the semiclassical
suppression exponent, one smears the tunneling probability over several
periods. The smeared probability does obey the scaling law (\ref{probab}).

The second feature of our system is as follows. We observe that the 
process under consideration is classically forbidden, and hence
exponentially suppressed, at arbitrary high energies. 
Our interest in this property is motivated
by the studies of similar processes in quantum field theory. As a
matter of fact, the exponential suppression at all energies is generic
for the field theoretical processes involving creation of some classical object 
(soliton, bubble of new phase or vacuum
configuration with different topology) in a collision of two highly
energetic quantum particles~\cite{Zakharov:1990xt,
Maggiore:1991kh,Voloshin:1993dk,Rubakov:1994hz,
Levkov:2004tf}. Moreover, it has been shown
recently~\cite{Levkov:2004tf} that the method of complex trajectories
predicts 
the suppression exponent $F$ of the above processes to attain 
its minimum at a certain ``optimal'' energy $E_o$,
above which $F$ stays
constant (this behavior of the suppression
exponent was conjectured earlier in
Refs.~\cite{Maggiore:1991kh,Voloshin:1993dk}).   
The optimal value  
$F(E_o)$ is determined by the complex--valued
classical solutions with particular properties, see
Refs~\cite{Levkov:2004tf}; these solutions are called
``real--time instantons''. 

One might question the
applicability of the method of complex trajectories for the 
description of the above phenomenon of optimal tunneling. Indeed, the
properties of processes at $E\approx E_o$ are in many respects
different from the 
well--known case of tunneling through a potential barrier. 
In this paper we provide
the evidence that the semiclassical method is applicable for
the description of dynamical tunneling independently of how high the
energy of the process is, or whether there exists a potential barrier
at all.

The model of this paper provides a particular example of
quantum mechanical system exhibiting the phenomenon of optimal
tunneling. Namely, the suppression exponent $F(E)$ depends on energy
$E$  in non-monotonic way, attaining a local minimum at
some energy $E_o$. We will find that the minimal value of the
suppression exponent is indeed given by the method of real-time
instantons. At higher
energies the function $F(E)$ grows to infinity\footnote{This is
  different from the case of collision--induced tunneling in field theory,
  where the suppression exponent stays constant at energies higher
  than $E_o$. The reason is that in the latter case 
at $E>E_o$ another tunneling
  mechanism, which is specific for the field theoretical setup, comes
  into play. Namely, above the optimal point the energy excess $(E-E_o)$
  is released by the emission of a few hard quantum particles, so that
  the tunneling transition effectively occurs at the optimal
  energy~\cite{Voloshin:1993dk,Levkov:2004tf}.}. 

The paper is organized as follows. In Sec.~\ref{sec:2} we describe the
model under consideration and introduce notations. In Sec.~\ref{sec:3}
the classical dynamics of the system is analyzed. The semiclassical
study of the classically forbidden reflections is performed
in Sec.~\ref{sec:4}. In Sec.~\ref{sec:5} we present
the results of the numerical integration of the Schr\"{o}dinger
equation and discuss their comparison with the semiclassical
results. Sec.~\ref{sec:6} contains summary and conclusions.

\section{Setup}
\label{sec:2}
\begin{figure}[htb]
  \begin{center}
\includegraphics[width=0.95\columnwidth]{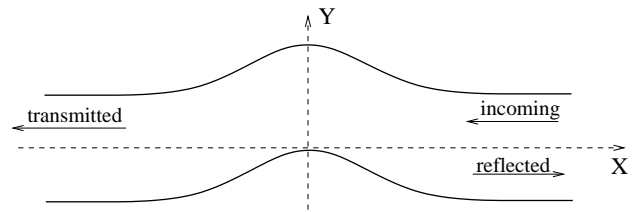}
    \end{center}
\caption{The equipotential contour ${\cal U}(X,Y) = {\cal E}$ for the
  waveguide model \eqref{ham1} and the directions of the incoming,
  reflected and transmitted fluxes of particles.}  
\label{fig:wav}
\end{figure}

Throughout the paper we illustrate our technique on a toy model
describing the evolution of a quantum particle in 
two-dimensional harmonic waveguide. Namely, we consider the case when the
motion of the particle is confined to the vicinity of a certain line
$Y = \frac{1}{g} 
a(g X)$ by the quadratic potential ($X$ and $Y$ stand for the
Cartesian coordinates of the particle). The equipotential contour
${\cal U}(X,Y)={\cal E}$ is shown in Fig.~\ref{fig:wav}; the
Hamiltonian is  
\be
\label{ham1}
{\cal H}=\frac{P_X^2+P_Y^2}{2m}+ {\cal U}(X,Y)\;,
\ee
where
\begin{align}
\label{ax}
&{\cal U}(X,Y) = \frac{m\omega^2}{2}
\left[Y-\frac{1}{g} a\!\left(g X\right)\right]^2\;, \\
\notag & a(x)=a_0\e^{-x^2/2}\;.
\end{align}
In the asymptotic regions $X\to\pm\infty$,
the variables of the model (\ref{ham1}) separate, and the
motion of the particle becomes trivial: 
oscillations in the $Y$-direction are accompanied by
the translatory motion along the $X$-axis.
The wriggle around the point $X\approx0$ introduces
nonlinear coupling between the degrees of freedom; we refer to
this  part of the configuration space as the interaction region.

In what
follows we use the system of units where
\be
\label{units}
\hbar=\omega=m=1\;.
\ee
The rescaling 
\begin{equation}
\label{eq:20}
{\boldsymbol X} = {\boldsymbol x}/g\;, \qquad
{\boldsymbol P} = {\boldsymbol p}g 
\end{equation}
brings the Hamiltonian (\ref{ham1}) into the form
\be
\label{ham2}
{\cal H}=\frac{g^2(p_x^2+p_y^2)}{2}
+\frac{1}{2g^2}(y-a(x))^2\;.
\ee
Equation \eqref{ham2} implies that the semiclassical regime in our
model occurs for $g^2\ll 1$.   
Apart from $g^2$, the only free parameter of the model~\eqref{ham2} is
$a_0$, which we set
\be
\label{oura0}
a_0=0.8\;.
\ee
This choice will be explained in Sec. \ref{sec:3}. To avoid confusion,
we remark that the rescaled coordinates $x$, $y$ will be used for the
semiclassical analysis of Secs.~\ref{sec:3},~\ref{sec:4} and
App.~\ref{app:A}, while the original ones ($X$ and $Y$) are exploited
in the quantum computations of Sec.~\ref{sec:5} and App.~\ref{app:B}. 

The process we want to investigate is the backward 
reflection of a particle coming
from the right, see Fig.~\ref{fig:wav}.  
Note that though we will refer to this process as over-barrier
reflection, there is actually no 
potential barrier in our system: the minimum of the potential is zero
in any transverse section of the waveguide.
The incoming quantum state $\ket{{\cal E},{\cal N}}$ is completely
determined by the total energy ${\cal E}$ and 
occupation number ${\cal N}$ of the $Y$-oscillator.
The quantity of interest is the total reflection 
coefficient for this state,
\begin{multline}
\label{rcoef}
{\cal P}({\cal E},{\cal N})=\lim_{t_f - t_i\to +\infty} 
\frac{1}{t_f-t_i} \times \\\times \sum_{f}
\big|\bra{f}\e^{-i{\cal H}(t_f-t_i)}\ket{{\cal E},{\cal N}}\big|^2\;,
\end{multline}
where $|f\rangle$ stands for the basis of reflected waves ($P_f > 0$)
  supported in the right asymptotic region, and the proper
  normalization of the incoming state has been chosen,
\be
\label{normal}
\braket{{\cal E},{\cal N}}{{\cal E}',{\cal N}'}
=2\pi\delta({\cal E}-{\cal E}')\delta_{{\cal NN}'}\;.
\ee
At some values of the initial--state parameters ${\cal E}$, ${\cal N}$  
the reflection process is classically forbidden. [In particular, one
  expects a particle with high enough translatory momentum $|P_i|$ to pass
  classically to the other side of the waveguide ending up in the 
asymptotic region $X\to -\infty$.] In this case the
reflection coefficient (\ref{rcoef}) is expected to 
obey the semiclassical scaling law (\ref{probab}) at $g^2\to 0$.
Below we concentrate on the calculation of the leading
suppression exponent $F$ as function of ${\cal E}$, ${\cal N}$.

We do not pretend to describe any concrete experimental situation by
the Hamiltonian  (\ref{ham1}); it is chosen as a convenient testing
ground of our semiclassical technique. The advantage of the 
scattering setup is an  unambiguous determination of
the initial and final states of the tunneling process, the latter  
determination being problematic in the case of
bounded motion~\cite{Wilkinson:Takada}. Moreover, the simple form of our
model makes it tractable both semiclassically and
by the exact quantum mechanical methods.
Note that a system similar to (\ref{ham1}) was considered in 
Ref.~\cite{Drew2005}.

\section{Classical reflections}
\label{sec:3} 
Let us start by considering the classical dynamics of the reflection
process. As we will see in Sec.~\ref{sec:4}, the analysis of the
classical dynamics is crucial for understanding the classically
forbidden  reflections. The results of the present Section will enable
us to identify the sets of the initial data $({\cal E},\; {\cal N})$  which
correspond to the classically allowed/forbidden reflections; for
brevity we refer to these sets as classically
allowed/classically forbidden regions. Note that these are the regions 
in the {\it initial data plane} $({\cal E}, \,{\cal N})$; they should
not be  confused with the classically allowed/forbidden regions in the
configuration space. The latter term is not used in this paper.

One notes that the action functional of the model (\ref{ham2}) has the
form 
\begin{equation}
\label{eq:21}
{\cal S}=S/g^2,
\end{equation}
where
\be
\label{clact}
S=\int \left(\frac{\dot x^2+\dot y^2}{2}-\frac{1}{2}
(y-a(x))^2\right) dt\;
\ee
does not contain the parameter $g$ at all. Hence, $g$  
drops out of the classical equations of motion.  While studying the
classical dynamics it is convenient to forget about $g$ and consider
the classical system defined by the rescaled action (\ref{clact}).

In contrast to quantum mechanics, where two quantum numbers ${\cal
  E}$, ${\cal N}$ determine the initial state completely, the
  classical evolution is specified by four initial conditions. One of these 
is physically irrelevant. It corresponds to the $x$ coordinate at the
  initial moment $t = t_i$, and can be absorbed by the appropriate
  shift of $t_i$. Note that, still, one 
should be careful to choose $x(t_i)$ far from the
interaction region; in numerical calculations of this Section 
  the value  
\bseq
\label{initial01}
\be
\label{initial0}
x(t_i)=10
\ee 
is used.
To keep contact with the quantum mechanical formulation
we choose the other two initial
data to be the total classical energy 
$E$ and the ``classical occupation number'' $N$, which is 
equal\footnote{Recall that in our units $\omega = 1$.} to the initial
classical energy of the transverse oscillations.  
This determines the initial velocity of the particle along the waveguide,
\be
\label{initial1}
\dot x(t_i)=-\sqrt{2(E-N)}\;.
\ee
\eseq
It is worth noting that Eq.~(\ref{eq:21}) implies the following
relation between the classical parameters and their quantum
counterparts, 
\be
\label{qcl}
E=g^2{\cal E}~,\qquad N=g^2{\cal N}\;.
\ee 
The last initial condition is the initial phase $\phi_0$ of 
$y$-oscillator. It parametrizes the initial position and velocity 
of the particle
in the transverse section of the waveguide:
\bseq
\label{initial23}
\begin{align}
\label{initial2}
&y(t_i)=\sqrt{2N}\cos\phi_0\;,\\
\label{initial3}
&\dot y(t_i)=-\sqrt{2N}\sin\phi_0\;.
\end{align}
\eseq
Clearly, a given point of the plane $(E,N)$ belongs to the classically
allowed region if classical reflection is possible for some value(s)
of $\phi_0$. Otherwise, we say that this point  lies in the
classically forbidden region.

At first glance it may seem that classical reflections are impossible
for any values of $E$, $N$ as there is no potential barrier to prevent
the classical particle from going into the left asymptotic region. 
Let us make sure that this is not the case by considering the
evolution at small $E$. In this regime the particle moves slowly along
the axis $y = a(x)$ of the waveguide performing small and
(relatively) rapid oscillations in the orthogonal direction. 
The frequency $\omega_\perp$ of the latter oscillations is determined by
the curvature of the transverse section of the potential
and thus depends on the position $x$ of the particle. 
It is straightforward to find  that   
$\omega_{\perp}(x)=\sqrt{1+(a'(x))^2}$.
The energy of the orthogonal 
oscillations $E_{\perp}$ divided by their frequency
is an adiabatic invariant,
\be
\label{adiab}
\frac{E_{\perp}}{\omega_{\perp}(x)}=const=N\;.
\ee 
On the other hand, the conservation of total energy yields,
\be
\label{energadiab}
E=\frac{v_{\parallel}^2}{2}+E_{\perp}=
\frac{v_{\parallel}^2}{2}+N\sqrt{1+(a'(x))^2}\;,
\ee
where $v_{\parallel}$ is the projection of the particle velocity onto
the axis of the
waveguide. Thus, the adiabatic motion in our waveguide is governed by
one--dimensional dynamics in the effective potential
\be
\label{Uadiab}
U(x)=N\sqrt{1+a_0^2x^2\e^{-x^2}}\;.
\ee
where the explicit expression (\ref{ax}) for the
function $a(x)$ has been used.
This picture is valid as long as $\dot\omega_\perp/\omega_\perp^2\ll 1$, 
which is satisfied for $E, N\ll 1$.

\begin{figure}[htb]
  \begin{center}
\includegraphics[width=0.9\columnwidth]{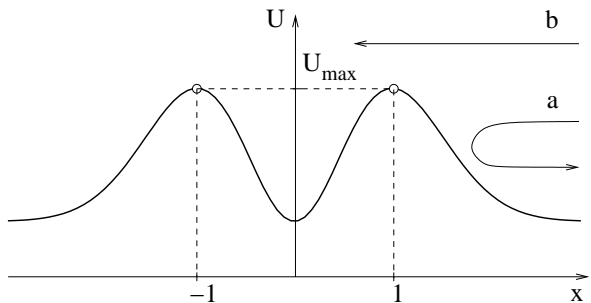}
    \end{center}
\caption{The effective potential for the motion in the adiabatic
  regime. The particle is reflected at $E<U_{max}$ 
  (case a), and is transmitted through the waveguide at
$E>U_{max}$ (case b).} 
\label{fig:Uadiab}
\end{figure}

The potential (\ref{Uadiab}) 
has the form of two symmetric humps with the maxima 
$U_{max}=N\sqrt{1+a_0^2\e^{-1}}$
situated at $x=\pm 1$, see Fig~\ref{fig:Uadiab}. 
Any particle coming from 
the right with $E<U_{max}$ gets reflected back; so, these
values of $E$, $N$ belong to the classically allowed region. In the
opposite case  $E>U_{max}$ the particle 
overcomes the effective potential which means that the reflection
process  is
classically forbidden. Thus, the 
line $E=N\sqrt{1+a_0^2\e^{-1}}$ is the boundary of the classically
allowed region at $E, N\ll 1$. 

When the values of the parameters 
$E$, $N$ approach the boundary of the classically allowed region
the particle spends more and more time around the tops of the effective
barrier $U(x)$. 
For the values precisely at this boundary there are
two unstable solutions $x=\pm 1$. In the  two-dimensional picture
these solutions correspond to periodic oscillations around the
fixed points  
$(x=\pm 1, y=a_0\e^{-1/2})$ 
on the axis of the waveguide. We will see that such 
unstable periodic solutions
exist beyond the adiabatic approximation
and play a key role in the semiclassical analysis, 
cf. Refs.~\cite{Onishi2003,Bezrukov:2003yf,Takahashi2003}. 
Borrowing the terminology from gauge theories
\cite{Klinkhamer:1984di} we call
them   
{\em excited sphalerons}\footnote{The word {\em sphaleron} is formed
from the Greek adjective
$\sigma\varphi\alpha\lambda\epsilon\rho o s$ meaning ``ready to
fall''.}. 
The sphaleron living at $x\approx 1$ ($x\approx-1$) will be referred to as
near (far) sphaleron according to its position relative 
to the right end of the waveguide.

\begin{figure}[t]
  \begin{center}
   \includegraphics[width=0.95\columnwidth]{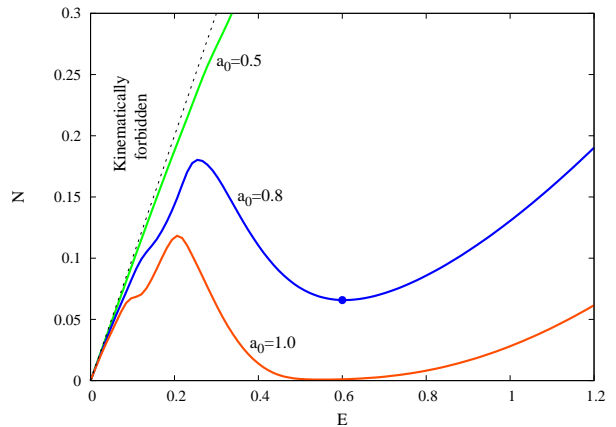}
    \end{center}
\caption{The boundaries $N_b(E)$ of the classically allowed regions
  plotted for $a_0 = 0.5;\, 0.8;\, 1.0$. The classical reflections
  are forbidden at $N<N_b(E)$. The points to the left 
  of the dashed line $E=N$ correspond to the kinematically forbidden
  initial conditions. }
\label{fig:1}
\end{figure}

To identify the boundary of the classically allowed region beyond the
adiabatic regime one resorts to numerical methods. We 
scan through the range of initial phases $0\leq \phi_0\leq 2\pi$ 
at fixed $E$, $N$ and look for the reflected trajectories. 
If such a trajectory is found at some $\phi_0$,
the point $(E,N)$ is identified as belonging to the classically
allowed region. Otherwise the point is attributed to the
classically forbidden domain. The results of these calculations are
presented in Fig.\ref{fig:1}.    
The boundaries $N_b(E)$ of the classically
allowed region are obtained for three different values of the
parameter $a_0$. The classical reflections are allowed for the initial
data above the boundaries, $N>N_b(E)$, and forbidden below them, 
$N<N_b(E)$.  
One observes that beyond the adiabatic regime
the form of the boundary $N_b(E)$ 
depends {\it qualitatively} on the value of $a_0$. 
At small $a_0$ it
monotonically increases.
At $a_0\approx 0.5$ a dip in the
curve $N_b(E)$ develops around $E\approx 0.6$. 
As $a_0$ grows further, this dip becomes lower and more pronounced,
until it touches the line $N=0$ at $a_0\approx 1$. At even larger
values of $a_0$ the boundary of the classically allowed region splits into two
disconnected parts and a range of energies around 
$E=0.6$ appears, where the classical reflections are allowed even for $N=0$.

Heuristically, 
one may envision that the form of the curve $N_b(E)$ reflects 
the behavior of the suppression exponent $F(E,N)$ in
the classically forbidden region. Indeed, the suppression exponent is
zero above the line $N = N_b(E)$.  As the value of $N$
gets decreased at fixed energy $E$, the function $F(E,N)$ starts
growing at $N = N_b(E)$. Thus, the deeper the point is in the classically
forbidden domain, the larger is $F$, and vice versa. 
According to this reasoning the dip in the curve $N_b(E)$ at 
$a_0>0.5$ implies that classically forbidden reflection are least
suppressed at energies $E\approx 0.6$. Moreover, at $a_0<1$ there is a
finite range of occupation numbers, 
$0<N<N_b(E=0.6)$, where the reflection process is
suppressed at all energies (except for the narrow band 
$N<E<N\sqrt{1+a_0^2\e^{-1}}$ corresponding to the adiabatic regime). 
For these values of $N$, the suppression exponent 
$F$ considered as function of $E$ is expected\footnote{We stress that
  the heuristic arguments about the behavior of the suppression
  exponent will be confirmed by the explicit semiclassical and 
  quantum mechanical calculations in the subsequent sections. } to have 
a (local) minimum in the vicinity of $E=0.6$. One of 
the purposes of this paper is
to test the method of complex trajectories in the regime when the
minimum of $F(E)$ exists; so, we concentrate on the case $a_0 = 0.8$.

\begin{figure}
\begin{center}
\includegraphics[width=0.4\textwidth]{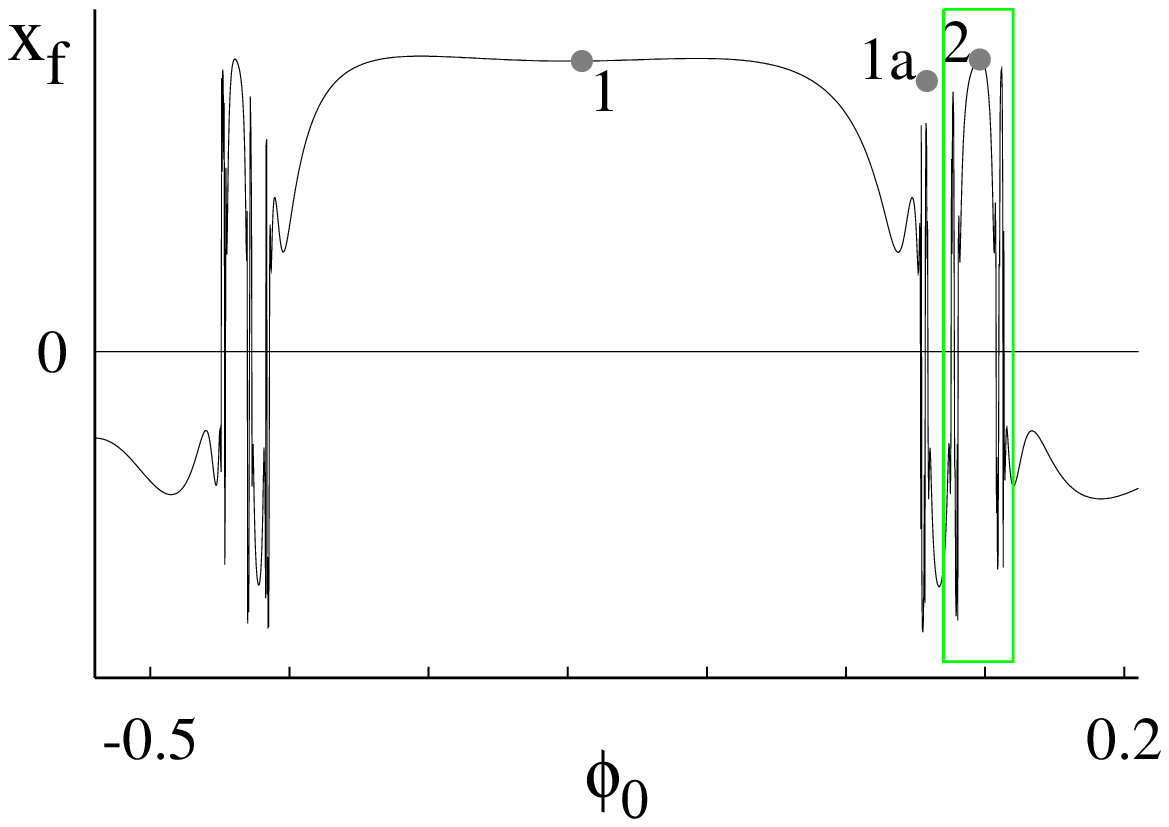}
\includegraphics[width=0.4\textwidth]{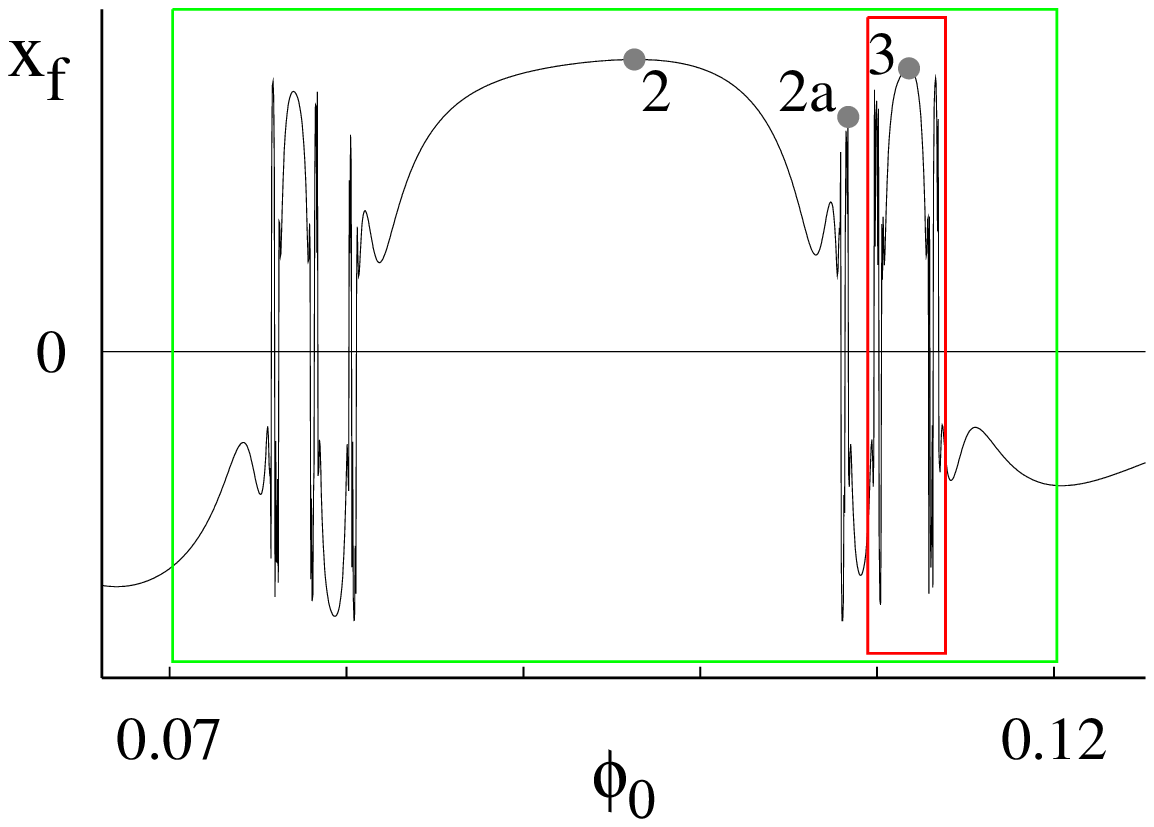}
\includegraphics[width=0.4\textwidth]{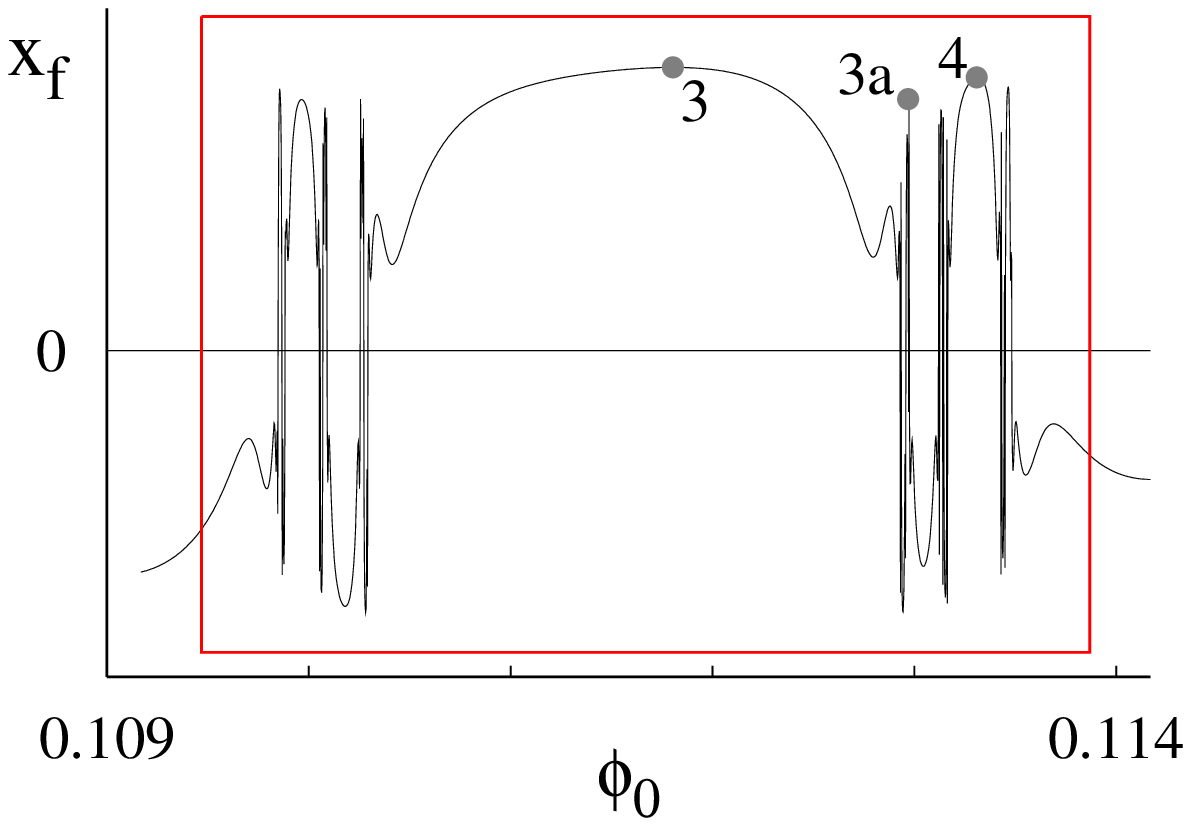}
\end{center}
\caption{The dependence of the final particle position on the value of
  the initial phase $\phi_0$ for $E=0.6$, $N=0.1$ ($x_f$ axis is not
  to scale). From top to bottom: scaling
  of the fine structures of the function $x_f(\phi_0)$ reveals
  self-similar behavior. The corresponding regions in different graphs
  are marked by the boxes of the same color. The
  trajectories corresponding to the marked points are shown in
  Fig.~\ref{fig:3}.}
\label{fig:2}
\end{figure}

Now, we are in a position to investigate the classical dynamics of the
system \eqref{clact} in detail. The observations we
make below are crucial for the subsequent study of the classically
forbidden reflections. One asks the following question. At given
$E,N$ belonging to the classically allowed region, there is a non-empty set 
${\cal R}_{E,N}$ of initial phases $\phi_0$, which give rise to
classical reflections. What is the structure of this set? 
To answer this question, we fix the initial
conditions (\ref{initial01}), (\ref{initial23}) at $t_i=0$ and integrate the equations of motion
until $t_f=200$ starting from different values of the initial phase $\phi_0$.
In this way, the dependence of $x_f\equiv x(t_f)$ on 
$\phi_0$ is obtained, see Fig.~\ref{fig:2}. 
The negative values of $x_f$ correspond to the classical
transmissions through the waveguide, while $x_f>0$ represent
reflections. Thus, 
\be
{\cal R}_{E,N}=\{\phi_0 | x_f(\phi_0)>0\}\;.
\ee 
Figure~\ref{fig:2} shows that the set ${\cal R}_{E,N}$ is not
connected: the intervals of 
phases corresponding to the reflected trajectories are intermixed with
those representing transmissions. Moreover, the scaling of the fine
structures of the function $x_f(\phi_0)$ reveals self-similar
behavior. One concludes that the set ${\cal R}_{E,N}$ consists 
of an infinite number of disconnected domains forming a fractal
structure. Such a complexity is a manifestation of irregular dynamics
inherent in our model; this feature is in sharp  contrast to the
situation one observes in  completely regular systems (see, e.g., the
model of Ref. \cite{Bonini:1999kj} where the analogous
set consists of a single
interval~\cite{Bezrukov:2003yf}).  

\begin{figure}[htb]
  \begin{center}
\includegraphics[width=0.7\columnwidth]{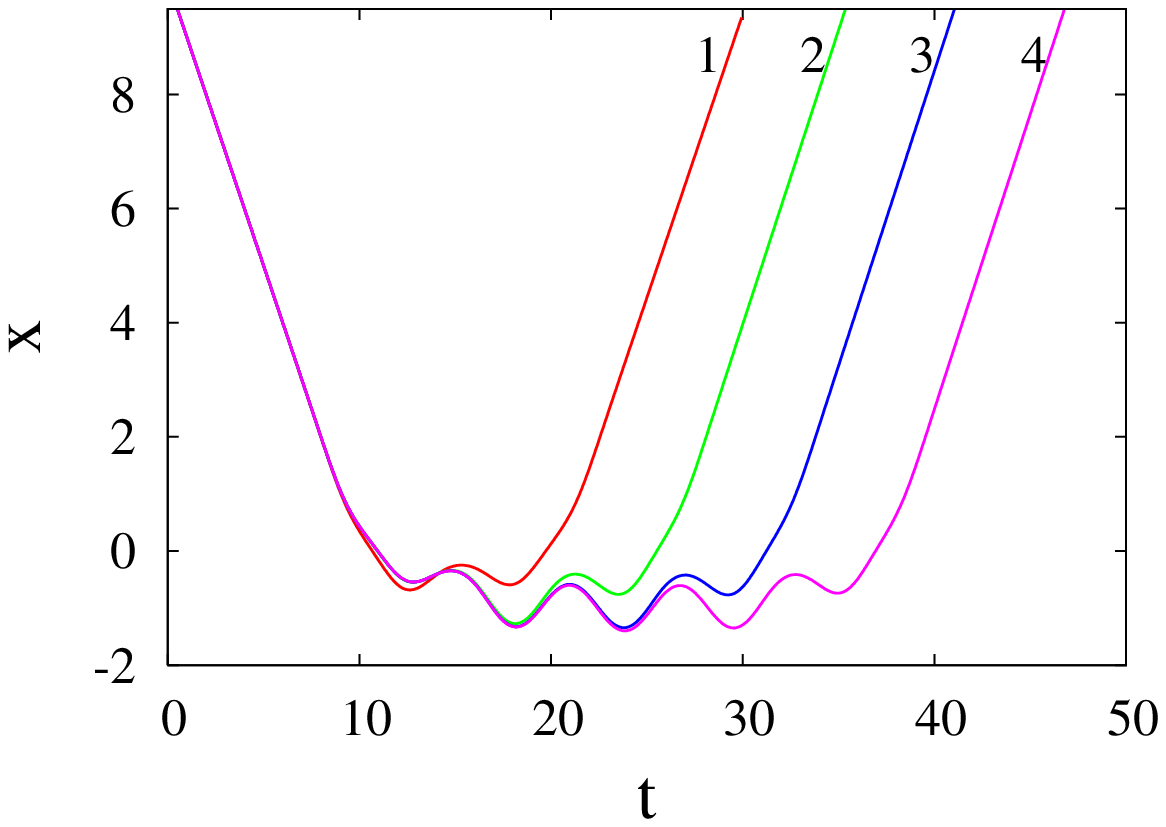}\\
~~~~~(a)\\
\includegraphics[width=0.7\columnwidth]{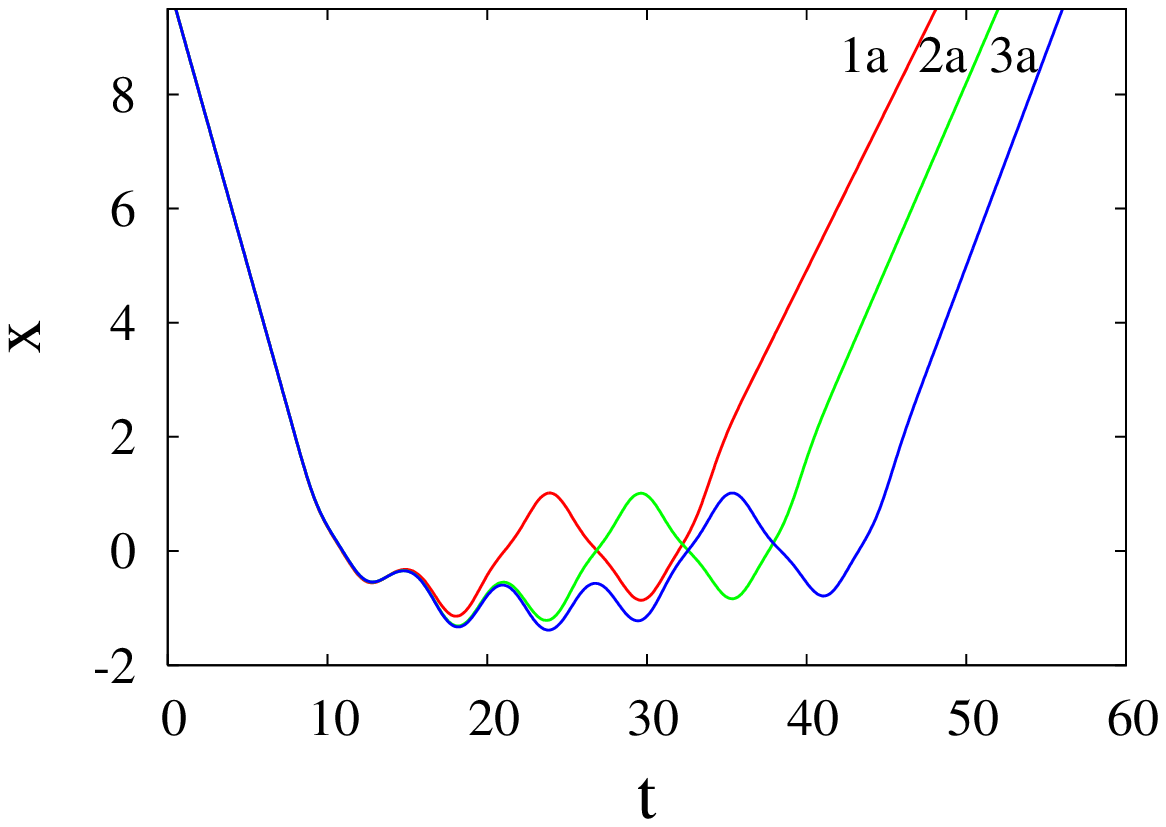}\\
~~~~~(b)
   \end{center}
\caption{The main (a) and secondary (b) sequences of
trajectories for $E=0.6$, $N=0.1$.
 The trajectories correspond to initial phases with the same marks as
 in Fig.\ref{fig:2}.}
\label{fig:3}
\end{figure}

To analyze the nature of irregular dynamics, we consider the 
trajectories generated by the initial phases which span various
connected intervals ${\cal R}_{\alpha; \, E,N} \subset {\cal
  R}_{E,N}$. All the classical 
trajectories from a given interval of phases display the same
qualitative properties; some features change discontinuously,
however, as one goes to another interval. 
Let us characterize each trajectory by its behavior in the interaction
region. To start with, one can distinguish a subset of intervals 
${\cal R}_{j;\, E,N} \subset {\cal R}_{E,N}$ corresponding to 
the trajectories which reach
the far sphaleron, perform several oscillations
there, and go out of the interaction region back to $x\to +\infty$.
We call this subset ``the main sequence''. 
The $x(t)$-dependence for the first four trajectories from the main
sequence is plotted in Fig.~\ref{fig:3}a, while
the corresponding values of the initial phase
are marked in Fig.~\ref{fig:2} by numbers (1 to 4).
Let us make the important observation:
all the plotted solutions leave the far sphaleron with
approximately the same oscillatory phase. In fact, we have found that
this is true for any reflected  trajectory. Here lies the root of the
disconnectedness of ${\cal R}_{E,N}$: one cannot transform one
reflected trajectory into another with a different number of
oscillations at the far sphaleron by continuously changing the initial
phase.  For a solution from the main sequence the index $j$
is equal to the number of oscillations at $x\approx -1$. 

The trajectories corresponding to the intervals
of ${\cal R}_{E,N}$ other than those from the main sequence display
more involved behavior. After oscillating at the far sphaleron they
move to the near sphaleron, oscillate an {\em integer}
number of periods on top of it, return to the far sphaleron, oscillate
there once again, etc. As an example we present several trajectories
from the next-to-the-main sequence (with only one return to the far
sphaleron) in Fig.~\ref{fig:3}b, they correspond to the initial phases
marked by 1a to 3a in Fig. \ref{fig:2}. We observed that the motion back
and forth between the sphalerons can be arbitrarily complicated giving
rise to the aforementioned fractal structure of  ${\cal
  R}_{E,N}$. 

\begin{figure}[htb]
  \begin{center}
\includegraphics[width=0.7\columnwidth]{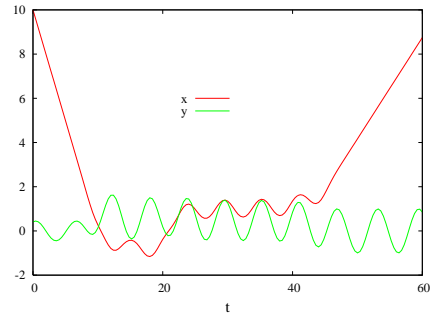}
    \end{center}
\caption{The classical trajectory with the value of the initial phase
  $\phi_0=-0.414$  close to the end--point, 
$\phi_0-\phi_{end} \sim 10^{-3}$, of the first
  interval of the main 
sequence; $E=0.6$, $N=0.1$. 
Before leaving the interaction region the trajectory gets stuck 
at the near sphaleron for a long time.} 
\label{fig:3.5}
\end{figure}

For the semiclassical analysis of the next Section it is important to 
know what happens with the classical reflected solution when
the initial phase $\phi_0$ approaches the end-point $\phi_{end}$ of an 
interval ${\cal R}_{\alpha;\, E,N}\subset {\cal
  R}_{E,N}$. Figure~\ref{fig:2} implies that at the 
ends of the interval  the corresponding trajectory gets stuck in the
interaction region. More precisely, one 
observes that as the value of $\phi_0$ approaches $\phi_{end}$, the
classical solution spends more and more time at $x \approx 1$ before
going away to infinity (see Fig.~\ref{fig:3.5}), so that  
the initial datum
$\phi_0=\phi_{end}$ gives rise to a trajectory which is neither
reflected, nor transmitted, but ends up at the near 
sphaleron. Such a trajectory is unstable: there exist arbitrarily small
perturbations which push it out of the interaction region 
to either end of
the waveguide. As the number of intervals (and hence of their
end-points) constituting ${\cal R}_{E,N}$ is infinite, the number of 
the above unstable trajectories is infinite, too.  
We will see in the next Section that the
above unstable solutions give rise to the 
tunneling trajectories; thus, the classically forbidden reflection
process of the next Section provides a particular example of
chaotic tunneling.

\begin{figure*}[t!]
  \begin{center}
\includegraphics[angle=-90,width=0.75\textwidth]{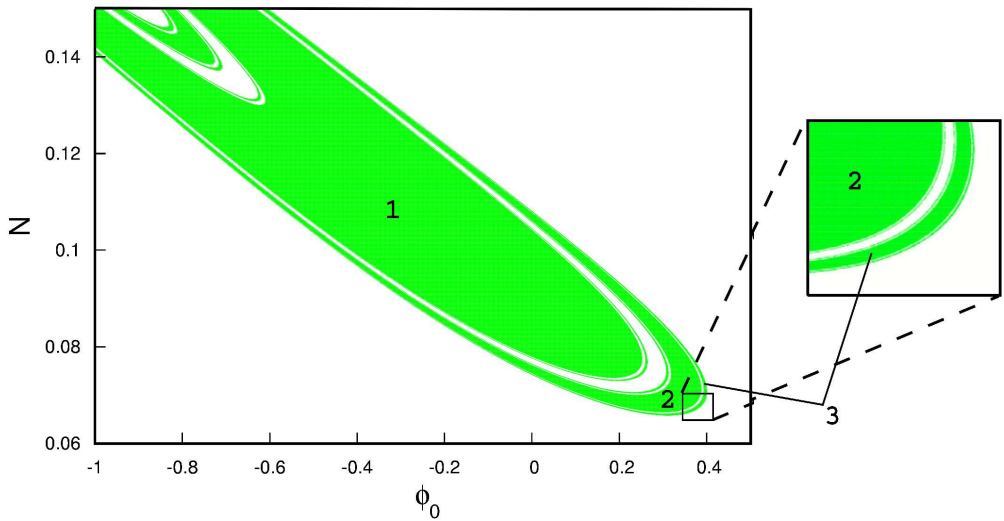}
    \end{center}
\caption{The set ${\cal R}_E$ of initial data leading to reflections at 
$E=0.6$ is painted in green. The inset shows the fine structure near
the boundary of ${\cal R}_E$. Three first domains of
the main sequence are marked by numbers.} 
\label{fig:Nphi}
\end{figure*}

More insight into the classical dynamics is gained by extending the
previous analysis to include the
variation of two initial conditions, $\phi_0$ and $N$, at fixed energy
$E$. One is again interested in the structure of the set ${\cal R}_E$ of
initial data $(\phi_0,N)$ which correspond to classical
reflections. This set is shown in Fig.~\ref{fig:Nphi} ($E=0.6$). 
As before, it consists of an infinite number of
disconnected domains ${\cal R}_{\alpha;\, E}$. Each of these domains
is characterized by the 
way in which the corresponding trajectories travel back and forth
between the sphalerons. The domains of the main sequence with 
$j=1,2,3$ are
clearly visible in Fig.~\ref{fig:Nphi} along with the secondary ones
which accompany them. Figure~\ref{fig:Nphi} enables one to understand
what happens with the classical reflected trajectories when the
occupation number $N$ approaches the boundary of the classically
allowed region from above, $N\to N_b(E)$; this corresponds
to moving to the lower boundary of the set ${\cal R}_E$,
$$
N_b(E)=\inf_{(\phi_0,N)\in {\cal R}_E} N\;.
$$
One notices that no matter how close $N$ is to this boundary,
the line $N=const$ always has intersections with some domains from the main
sequence. In other words, the boundary of the classically
allowed region is the accumulation point of the main sequence of
domains. On the other hand, as $N$ tends to $N_b(E)$, the intersection
of the line $N=const$ 
with certain domains of ${\cal R}_E$ disappears. This means
disappearance of certain types of trajectories at $N\to
N_b(E)$. Concentrating on the main sequence, one observes that the
remaining
trajectories are those with large indices $j$. As the latter are equal to
the number of oscillations at the far
sphaleron, one concludes that the
closer the point $(E,N)$ is to the boundary of the classically allowed
region, the longer  the corresponding trajectories get stuck at the far
sphaleron\footnote{This property should not be confused with the
  property described in the previous paragraph, where it was pointed
  out that the trajectories get stuck at the {\em near} sphaleron when the
initial data approach the boundary of {\em a single domain} 
${\cal R}_{\alpha;\,E}$ of 
${\cal R}_E$.}. 

We conclude this section with remarks
on the implications of the above classical picture 
for the semiclassical description of tunneling.
It will be shown in Sec.~\ref{sec:4} that each  
isolated domain of the set ${\cal R}_E$ continues into  a
branch of tunneling trajectories. Thus, the number of tunneling
solutions at given $E$, $N<N_b(E)$ is infinite; each of these
solutions is associated with a certain domain  ${\cal R}_{\alpha ;\,
  E}$. In particular,   
the fractal structure of ${\cal R}_E$ is inherited by the collection
of tunneling paths. The above property is very different from that
of the models with completely regular classical dynamics 
\cite{Bezrukov:2003yf} where the
tunneling solution is unique. On the other hand,
an infinite number of tunneling trajectories seems to be
generically inherent in chaotic systems 
\cite{Shudo1995_1}.   
According to the general rules, the suppression exponent $F$ in
Eq.~(\ref{probab}) is equal to the lower bound 
of the suppressions calculated on all the tunneling solutions,
\be
\label{Finf}
F(E,N)=\inf_{\alpha} F_\alpha(E,N)\;,
\ee
where the index $\alpha$ marks the  
solutions. At first 
glance it is not clear how the lower bound (\ref{Finf})  
can be found: one is
unable to compute the suppression exponents of an infinite set of
trajectories. Two observations which greatly simplify the problem are
as follows. First, it will be shown below that the nearer the domain 
 ${\cal R}_{\alpha;\, E}$ is to
the boundary of the classically allowed region, the smaller is the
suppression exponent of the tunneling solution associated with it.
Second, the domains of the main sequence accumulate
near the boundary of the classically allowed region.
This implies that the suppression exponents $F_j$
of the tunneling trajectories from the main sequence\footnote{These
  are the ones originated from the main sequence ${\cal R}_{j;\, E}$
  of domains. } decrease
with the index
$j$. Moreover, for any tunneling
path there exist tunneling solutions from the main sequence 
with the same $E$, $N$ and  smaller value of the suppression
exponent. Thus, in order to calculate the
lower bound (\ref{Finf}) it is sufficient to
consider the trajectories from the main sequence, compute
their suppression exponents $F_j$, and take the limit,
\be
\label{Flim}
F(E,N)=\lim_{j\to\infty}F_j(E,N)\;.
\ee   
This tactics is implemented in the next section.

\section{Semiclassical study}
\label{sec:4}
We now formulate the boundary value problem for the tunneling
trajectories. The probability (\ref{rcoef}) of over-barrier reflection 
is described by the complexified solution to the classical
equations of motion 
\bseq
\label{bvp}
\be
\label{bvp0}
\frac{\delta S}{\delta x(t)}=\frac{\delta S}{\delta y(t)}=0
\ee
obeying the following boundary conditions,
\begin{align}
\label{bvp1}
&\frac{{\dot x}^2}{2}+N=E~,~~~
\frac{{\dot y}^2}{2}+\frac{y^2}{2}=N, &\mathrm{at}~~~~~t\to
-\infty\;,\\
\label{bvp2}
&\Im x~,~~\Im y \to 0,
&\mathrm{at}~~~~~ t\to +\infty\;.
\end{align}
\eseq
Here $E$, $N$ stand for the (rescaled, see Eqs. (\ref{qcl})) energy and initial
occupation number.\footnote{Note that the boundary value problem
  (\ref{bvp}) is invariant with respect to time translations: if
  $x(t)$, $y(t)$ is a solution to Eqs. (\ref{bvp}), then $x(t+\tau)$,
  $y(t+\tau)$ is also solution for any $\tau \in {\mathbb{R}}$. We
  fix this ambiguity by requiring $\mathrm{Re}\, x(t_i) = 10$
  (cf. Eq. (\ref{initial0})).} 
The initial conditions (\ref{bvp1}) at
$t\to -\infty$ can be cast into the form
(\ref{initial1}), (\ref{initial23}), where the initial phase $\phi_0$
is now allowed to
take complex values. Similarly,
$x(t_i)$ is also complex, in general. 
The boundary conditions (\ref{bvp1}), (\ref{bvp2}) have clear physical
interpretation. Equations~(\ref{bvp1}) fix the quantum
numbers ${\cal E}$, ${\cal N}$ of the initial quantum state. 
Due to the uncertainty principle this makes the two conjugate
coordinates, $\phi_0$ and $x(t_i)$, maximally
indeterminate. Accordingly, in the
semiclassical picture they  become
complex--valued. On the other hand, 
the quantum numbers of the final states are
not fixed, and Eqs.~(\ref{bvp2}) imply 
that the particle comes out in  a classical state with real
coordinates and momenta.

Note that, generically, the tunneling solution is defined along a
certain  contour in the complex time plane. In our case, however, the
contour is trivial: it runs along the real
time-axis. 

It is useful to parametrize the imaginary
parts of $x(t_i)$ and $\phi_0$ as follows,
\begin{align}
\label{bvp3}
&2\Im x(t_i)=-\dot x(t_i) T\;,\\
\label{bvp4}
&2\Im \phi_0=-T-\theta\;,
\end{align}
where $T$ and $\theta$ are real parameters. Then the suppression
exponent of a given complex trajectory is
\be
\label{suppr1}
F_\alpha=2\Im \tilde S_{\alpha}-ET_\alpha-N\theta_\alpha\;,
\ee
where the two last terms result from 
the non--trivial initial
state of the process, while
\be
\label{suppr2}
\tilde S_\alpha=\frac12 \int dt \left(-x\ddot x
-y\ddot y-(y-a(x))^2\right)
\ee
is the classical action integrated by parts. The subscript $\alpha$ is
introduced to remind that there may exist several tunneling 
solutions with given $E$, $N$.
 
We do not present the derivation of the above boundary value
problem. The logic is completely analogous to that of
Refs.~\cite{Bonini:1999kj,Bezrukov:2003yf,prefactor,wigles}, and an
interested reader is referred to these papers. The field theory
analog of the problem (\ref{bvp}) was first
introduced in Ref.~\cite{Rubakov:1992ec}.  

It is important to remark that the problem
(\ref{bvp}) does not guarantee per se that its solutions 
describe reflections. 
To ensure that this is the case, one supplements
Eqs. (\ref{bvp}) with the condition
\begin{equation}
\label{reflect}
\Re x\to +\infty~~~~~~~\mathrm{at}~~~t\to +\infty\;. 
\end{equation}
Below we find solutions 
which satisfy this requirement by using the
$\epsilon$--regularization method of
Refs.~\cite{Bezrukov:2003yf}. 

Let us explain the physical meaning of the initial-state parameters $T$,
$\theta$. One can prove
the relations (see, e.g.,
Refs. \cite{Bezrukov:2003yf,wigles})
\begin{equation}
\label{eq:22}
T = -\frac{\partial}{\partial E} F(E,N) \;, \qquad \qquad 
\theta = -\frac{\partial}{\partial N} F(E,N)\;,
\end{equation}
which imply that  $T$ and $\theta$ are the
derivatives of the 
suppression exponent with respect to energy and initial oscillator
excitation number. One notices that  $T$, $\theta$
can be used instead of $E$, $N$ to parametrize the tunneling
paths. Then, the solutions with $T=0$ correspond to the extrema of the
suppression exponent with respect to energy. 
These solutions are called ``real--time instantons'', 
they can be found directly 
using the method of Ref.~\cite{Levkov:2004tf}. 
Calculating the value of the functional (\ref{suppr1}) on
them, 
one obtains the extremal (notably, 
minimal) values of the suppression exponent at $N = const$. The method
of real--time instantons is important in field
theory~\cite{Levkov:2004tf}, where it enables one to
calculate the minimal suppression of the collision--induced
tunneling. One of the purposes of this paper is to check the above method 
by the explicit comparison with the exact quantum
mechanical results. Accordingly, we pay specific attention to the
region around the minimum of the suppression exponent $F(E)$. 

We solve the boundary value problem (\ref{bvp}) numerically with
the deformation procedure. Namely, the solution with energy 
$E+\Delta E$ and oscillator excitation number $N + \Delta N$ is found
by the iterative Newton--Raphson method~\cite{NumericalRecipes}
starting from the solution at  $E$, $N$, which serves as the
zeroth--order approximation. In this way, an entire branch of  
tunneling trajectories can be obtained starting from a single
solution and 
walking in small steps in $E$, $N$. The details of our numerical
technique can be found in Refs.~\cite{Bonini:1999kj},
see Refs.~\cite{Kuznetsov:1997az,Bezrukov:2003er} for
the applications in field theory. 

The only non-trivial task of the above approach is to find the input
for the first deformation. The idea we put forward in this paper is to
start the procedure from the classically allowed region and
arrive at the tunneling solutions by changing  the
values of $E$, $N$ in small steps. [Note that the approach of
  obtaining complex trajectories from the real ones was also
  implemented in a different context in Ref.\cite{Xavier1996}.]  
One begins with the observation that the real
classical trajectories 
satisfy\footnote{Their suppression is obviously zero.} 
Eqs. (\ref{bvp}). Naively, one takes a trajectory with 
$E$, $N$ from the
classically allowed region, and decreases the value of $N$ with the 
hope to
get the correct tunneling solution at $N < N_b(E)$. However, 
serious obstacles arise on the way. 
First, the classical solution is not
unique at given $E$, $N$; the degeneracy is parametrized by the
initial phase $\phi_0\in {\cal R}_{E,N}$. 
This makes the numerical implementation of the deformation procedure
problematic. 
Second, suppose one takes the
initial data $(N,\phi_0)$ belonging to some connected domain
${\cal R}_{\alpha;\, E} \subset {\cal R}_E$ and decreases $N$. It was observed in
Sec.~\ref{sec:3} that, as the value of $N$ approaches the boundary $N
= N_{b,\alpha}(E)$ of ${\cal R}_{\alpha;\, E}$ from above, the reflected
classical solution spends more and more time in the interaction
region. At $N = N_{b,\alpha}(E)$ it gets stuck at $x\approx 1$
forever, merging at this point with the classical solutions, which
represent the transmissions of the particle through the waveguide at $N <
N_{b,\alpha}(E)$. Here lies the
worst obstruction: the transmitted classical trajectories obviously solve the
boundary value problem (\ref{bvp}), so that the deformation
procedure which starts from the classically allowed region will
produce them at $N < N_{b,\alpha}$, instead of the correct complex
reflected solutions. One 
concludes that the condition (\ref{reflect}) should be somehow
incorporated explicitly into the boundary value problem. 

A method which automatically fixes the asymptotic (\ref{reflect}) of the
solution is proposed in
Refs.~\cite{Bezrukov:2003yf,prefactor}. It is called 
$\epsilon$--regularization. 
Here we briefly describe this method concentrating on its application
to the problem at hand. A more detailed description of the technique
can be found in
Refs. \cite{Bezrukov:2003yf,prefactor}. One replaces
the action of the system in Eqs. (\ref{bvp}), (\ref{suppr1}) with the
modified action 
\be
\label{Seps}
S_{\epsilon}[{\boldsymbol x}]=
S[{\boldsymbol x}]+i\epsilon T_{int}[{\boldsymbol x}]\;.
\ee
Here $\epsilon$ is small and positive, while the functional $T_{int}$
measures the time the particle spends in the interaction region. The
simplest choice is
\be
\label{Tint}
T_{int}=\int dt f(x(t),y(t))\;,
\ee
where the function $f$ is real and
positive at $x,\,y\in {\mathbb{R}}$, and is localized in the
interaction region. Otherwise the choice of $f$ is arbitrary: the
final result is recovered in the limit $\epsilon \to +0$ and does not
depend on the particular form of this function. We use 
\be
\label{f}
f=\tilde\theta (1-x)\tilde\theta (x-1)\;,
\ee
where
\be
\label{tildetheta}
\tilde\theta (x)=\frac{1}{1+\e^{-2x-x^3}}
\ee
is the smeared $\theta$--function. Note that $f$ peaks at the near
sphaleron; this choice will be explained shortly. 

One notices two important changes that the substitution \eqref{Seps}
brings into the boundary value problem
(\ref{bvp}). First, the degeneracy of the classical solutions is
removed at $\epsilon\ne 0$. Indeed, any unperturbed 
classical trajectory extremizes
the original action functional $S[{{\boldsymbol x}}]$, so that at fixed
$E$, $N$ and $\epsilon=0$ there exists a valley of extrema
parametrized by $\phi_0$. The functional $T_{int}[{\boldsymbol x}]$,
however, discriminates among all these extrema and 
lifts the valley. Correspondingly, at small $\epsilon>0$ the
extrema  of the regularized action $S_{\epsilon}[{\boldsymbol x}]$ are
close to the classical reflected solutions with $\partial
T_{int}/\partial\phi_0 = 0$  
(see Refs.~\cite{Bezrukov:2003yf} for the detailed
discussion). From the physical viewpoint this can be understood as
follows. In the   
$\epsilon$--regularized case the suppression, Eq. (\ref{suppr1}),  of
the trajectories in the classically allowed region is not precisely
zero because of the complex term $i\epsilon T_{int}$ in Eq. (\ref{Seps}).
The suppression is minimized on
the real classical trajectories corresponding to the minima of
$T_{int}$; therefore, the
solutions to Eqs. (\ref{bvp}), (\ref{Seps}), which by construction 
correspond to the least
suppressed reflections, are close to the above classical
trajectories.

In practice, one starts with the function
$T_{int}(\phi_0)$ representing the values of $T_{int}$ 
on the classical trajectories with fixed $E$, $N$ and $\epsilon=0$, finds
the extrema of this function, and uses the corresponding 
trajectories as the zeroth--order approximation 
to the solutions of Eqs. (\ref{bvp}), (\ref{Seps}) with small,
but non-zero $\epsilon$. 

The second useful property of the $\epsilon$--regularization comes
about as one tries to decrease the value of the initial
oscillator excitation number, and cross the boundary
$N=N_{b,\alpha}(E)$ of a given classically allowed domain. One discovers 
that at $\epsilon\ne 0$ each  reflected trajectory at
$N>N_{b,\alpha}(E)$ is smoothly 
connected with the complex reflected solution at $N <
N_{b,\alpha}(E)$. The reason, again, lies in the additional
suppression caused by the 
term $i\epsilon T_{int}$ in the action. 
Although the reflected trajectories get only slightly perturbed at
small $\epsilon$, the solutions ending up in the 
interaction region at $t\to +\infty$ change drastically. 
Indeed, the functional
$T_{int}[{\boldsymbol x}]$ diverges on the latter trajectories giving rise to
the infinite suppression. The immediate consequence is that
the paths which get stuck in the interaction region are excluded
from the set of solutions to the boundary value problem (\ref{bvp}),
(\ref{Seps}). Now, the solutions to Eqs.~(\ref{bvp}),
(\ref{Seps}) cannot change their asymptotic (\ref{reflect}): 
starting from the classical
reflected solution and decreasing the value of the initial excitation
number $N$ in small steps, one leaves the classically allowed region
and obtains the correct complex reflected trajectories.

Let us briefly comment on the choice 
of the contour in the complex time plane
which carries the tunneling solutions. Obviously, the
classical trajectories, which are the starting point of our
deformation procedure, run along the real time axis. Generically, 
while moving into the
classically forbidden region, one may need 
to deform the time
contour in order to avoid the singularities of the solution, see
Refs.~\cite{Bezrukov:2003yf}. In the actual
calculations of this paper we did not encounter such a
necessity. The time contour was always kept coincident with the
real axis.

\begin{figure}[t!]
\begin{center}
\includegraphics[width=0.7\columnwidth]{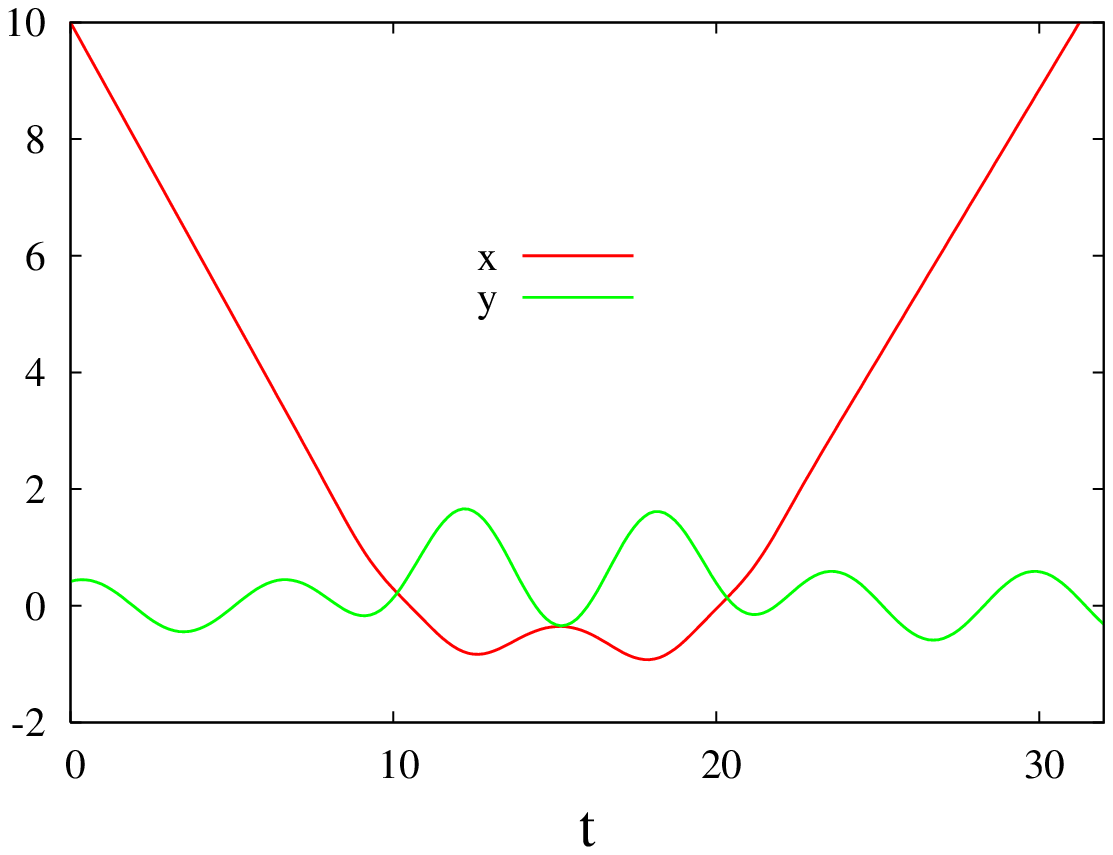}\\
~~~~(a)\\
\includegraphics[width=0.7\columnwidth]{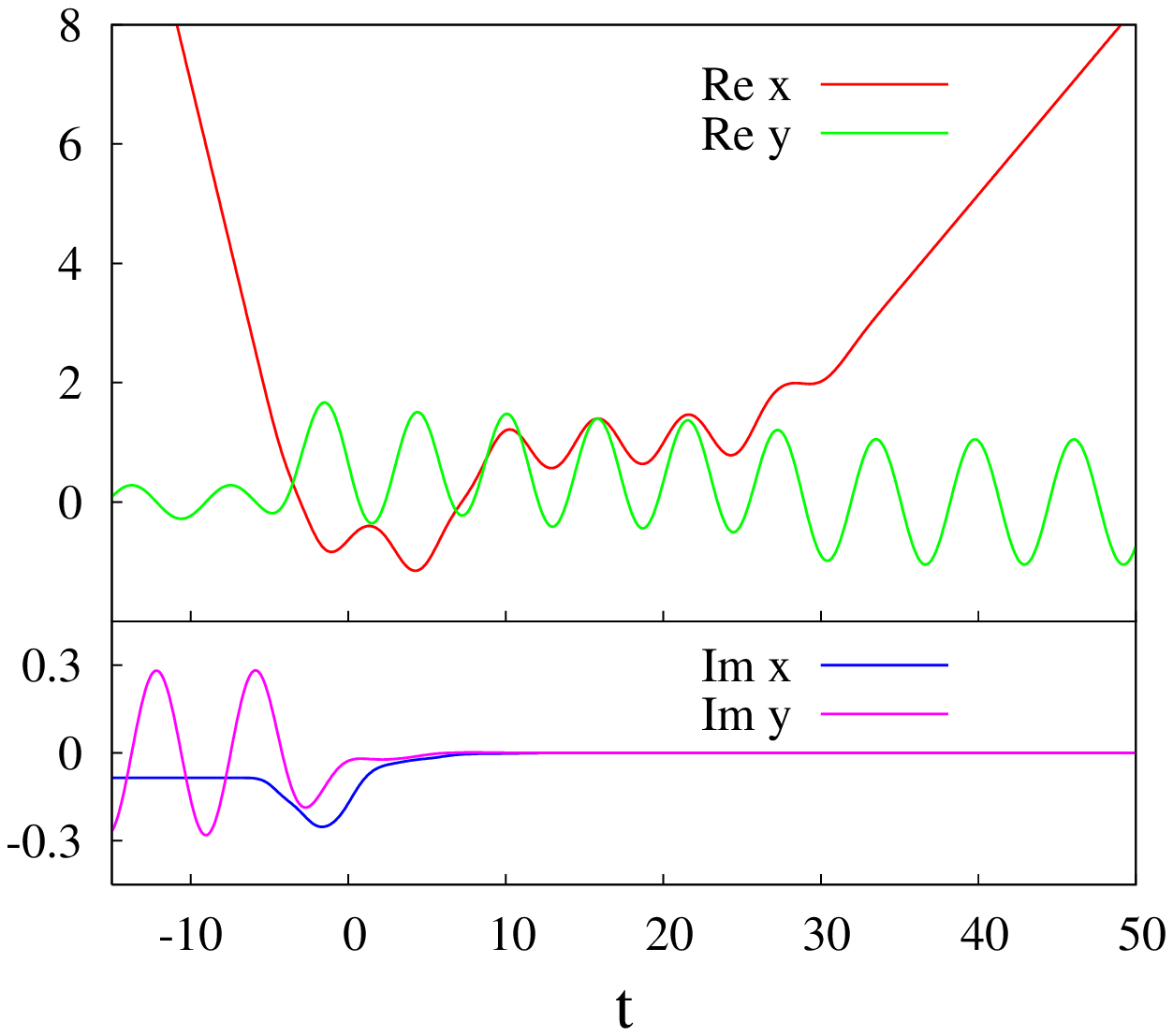}\\
~~~~(b)\\
\includegraphics[width=0.7\columnwidth]{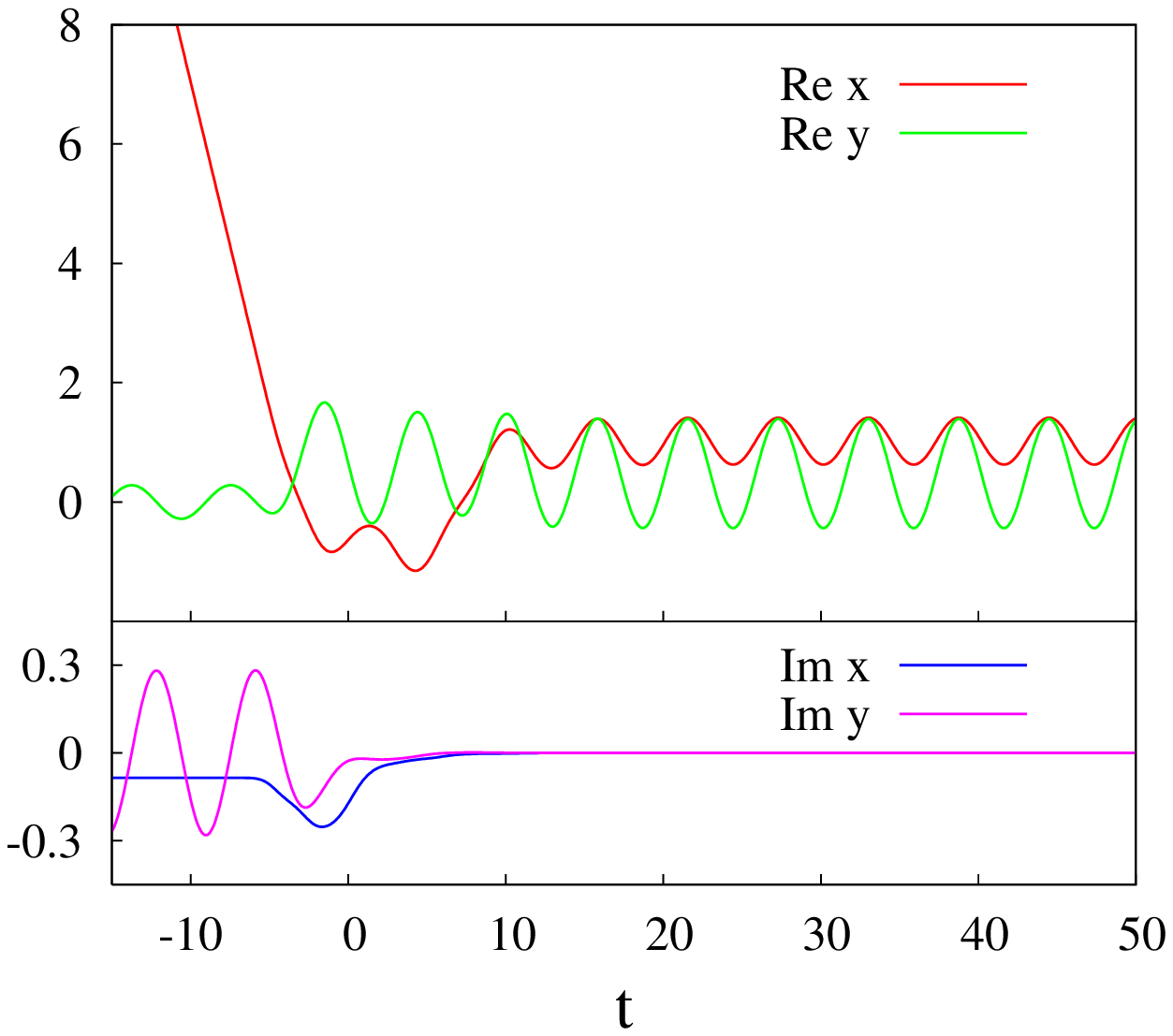}\\
~~~~(c)
\end{center}
\caption{\label{fig:walk_example1}
(a) The reflected classical trajectory from the first domain
  of the main sequence, corresponding to the minimum of
  $T_{int}(\phi_0)$ ($E=0.6$; $N=0.1$; $\phi_0=-0.366$). (b) The
  regularized tunneling solution ($E=0.6$, $N=0$, $\epsilon =
  10^{-6}$) descended from the above classical trajectory. (c) The
  limit $\epsilon\to 0$ of the tunneling solution.}
\end{figure}

An example of a regularized tunneling solution deep
inside the classically forbidden region ($E=0.6$, $N=0$) is shown in
Fig.~\ref{fig:walk_example1}b. 
The depicted solution descends from
the classical reflected trajectory with $E=0.6$, $N=0.1$, $\phi_0=-0.366$
(see Fig.~\ref{fig:walk_example1}a), which (locally) 
minimizes the function $T_{int}(\phi_0)$; 
the regularization $\epsilon=10^{-6}$ was switched on and $N$ was 
gradually decreased by small steps. 
Note that the solution in Fig.~\ref{fig:walk_example1}b is
genuinely complex and satisfies the requirement (\ref{reflect}). 

The solutions to the original boundary value problem are recovered in
the limit $\epsilon \to +0$; the suppression exponents of the 
unperturbed trajectories are
\begin{equation}
F_{\alpha}(E,N)=\lim_{\epsilon\to +0} F_{\alpha,\epsilon}(E,N)\;.
\end{equation}
Practically, the removal of the 
regularization is carried out by taking $\epsilon$ 
to be small enough, $\epsilon\lsim 10^{-6}$. At these $\epsilon$,
the values of the suppression exponents $F_{\alpha,\epsilon}$ 
stabilize at the level of accuracy $O(10^{-5})$ which is sufficient
for our purposes.

As one tries to take the limit $\epsilon\to +0$ of the tunneling
trajectory itself, a surprise comes about. Namely, as $\epsilon$ gets
decreased, the regularized tunneling 
trajectories deep inside the classically forbidden region stay longer
and longer at finite $x$, so that the trajectories with
$\epsilon=0$ do not escape to infinity, but end up on an unstable
solution -- the near sphaleron -- living at $x\approx 1$. [An example
of unperturbed tunneling trajectory ($E=0.6$, $N=0$, $\epsilon=0$)
is shown in Fig. \ref{fig:walk_example1}c.] 
Strictly speaking, the solutions at $\epsilon=0$ do not describe
reflection, but rather creation of the unstable state,
the sphaleron. Nevertheless, their suppression exponents are
relevant for tunneling as the latter state 
decays\footnote{Classically, the particle sits at the sphaleron
  for an infinite period of time. However, quantum
  fluctuations lead to the decay of this state with the carachteristic
  time of order~~$\ln{g}$.} 
into the asymptotic
region $x\to +\infty$ without exponential suppression. 
The mechanism of dynamical tunneling via creation of unstable periodic
orbits was discovered independently 
in Refs.~\cite{Bezrukov:2003yf} and Refs.~\cite{Takahashi2003}.

We remark that {\it all} unregularized 
tunneling trajectories in the model
under consideration tend to the unstable sphaleron solution at $t\to
+\infty$, and thus turn out to be unstable themselves. 
Straightforward numerical methods are inappropriate for finding such
trajectories \cite{Bonini:1999kj}, while
$\epsilon$--regularization provides a universal method for treating
such instabilities\footnote{An alternative would be to change the boundary
conditions (\ref{bvp2}), see App.~\ref{app:A}.}. 
 
Let us now discuss the structure of tunneling solutions. Following the
above strategy, one starts with the classical reflected solutions.
It is straightforward to see that 
the function $T_{int}(\phi_0)$ has at least one
extremum in every connected interval ${\cal R}_{\alpha;\, E,N}$ 
in the
set of reflection phases ${\cal R}_{E,N}$. Indeed, the classical
reflected trajectories with $\phi_0\in {\cal R}_{\alpha;\, E,N}$ spend
finite time in the interaction region and smoothly depend on
the initial data, producing the smooth function $T_{int}(\phi_0)$.
On the other hand, the classical trajectories spend more and more time
at the near sphaleron as $\phi_0$ approaches the end-points of ${\cal
  R}_{\alpha;\, E,N}$ (see Sec.~\ref{sec:3}). Thus, $T_{int}(\phi_0)$
tends to infinity at the end points of this interval, necessarily
attaining a minimum somewhere in between\footnote{In general, there
  might be several extrema of $T_{int}$ inside each connected interval
${\cal R}_{\alpha;\, E,N}$, and one should consider solutions
  corresponding to each of these extrema. In our case the minimum is
  unique. }. [Note that
  the specific form of the functional $T_{int}$ (function $f$ in
  Eq. (\ref{Tint}) peaks at $x\approx 1$) has been chosen in
  accordance with the tendency of classical trajectories to 
get stuck at the near sphaleron.]
Thus, starting from each interval ${\cal R}_{\alpha;\,E,N}$, one
obtains one branch of solutions to the $\epsilon$-regularized
problem. Note that, as ${\cal R}_{E,N}$ is the section of the
set ${\cal R}_E$ by the line $N=const$, the domains of 
${\cal R}_E$ may correspond to two disconnected intervals of ${\cal
  R}_{E,N}$, see Fig.~\ref{fig:Nphi}. When $N$ gets decreased, 
these intervals merge. Below we assume that the starting classical
solution of the type $\alpha$ has been taken at small enough $N$,
such that the domain ${\cal R}_{\alpha;\,E}$ gives rise to a single 
interval of phases\footnote{One legitimately asks what happens if the
starting classical solution is taken at large enough $N$, where two
different intervals of phases correspond to one and the same
domain of ${\cal R}_{E}$. Clearly, in this case  the above
deformation procedure will produce two different 
solutions to Eqs. (\ref{bvp}), (\ref{Seps}). We have observed,
however, that as the value of $N$ gets decreased below the point
where the two intervals merge, 
the corresponding solutions become identical.}  
${\cal
  R}_{\alpha;\, E,N}$. Then, the fractal structure of ${\cal
  R}_E$ is inherited by the complex tunneling paths: the distinct
branches of complex trajectories are in one--to--one
correspondence with the connected domains of the set ${\cal R}_E$.

Now, we can give 
a heuristic justification of the claim made in the end of
Sec.~\ref{sec:3}: the nearer the domain 
${\cal R}_{\alpha;\, E}\subset {\cal R}_E$ lies
to the boundary of the classically forbidden region, the less
suppressed is the corresponding tunneling trajectory. Let us
consider two branches of solutions stemming from the domains
1 and 2 in Fig.~\ref{fig:Nphi}. The line $N=0.07$ intersects with the
domain 2, so that the point $E=0.6$, $N=0.07$ belongs to the classically
allowed region, and the suppression of the  solution 2 vanishes in the 
limit $\epsilon\to 0$. On the other hand, the line $N = 0.07$ does not
cross the domain 1. Thus, at $E=0.6$, $N=0.07$ 
the corresponding solution to Eqs. (\ref{bvp}), (\ref{Seps}) is
genuinely complex and has non-zero
suppression even at  $\epsilon\to 0$. Suppose that one decreases $N$
and enters the classically forbidden region. At some point the
solution of the type 2 also becomes  classically forbidden. It is
clear that
at least for some range of $N$, its suppression remains
weaker than that of the solution 1. This suggests, though does not
guarantee, the same hierarchy of suppressions inside the entire
classically forbidden region.
Below, we check the conjectured hierarchy 
explicitly.

Let us discuss in detail the case $N=0$ (no transverse oscillations
in the initial state) corresponding to the extreme values 
of parameters inside the forbidden region; all the qualitative
features are the same for other values of $N$ as well. 
Our strategy is to find the limit 
(\ref{Flim}) of the main sequence of suppressions $F_j$, and check
that this limit represents the lower bound of the suppression
exponents of all the tunneling solutions.  
\begin{figure}[!ht]
\centerline{\includegraphics[width=0.68\columnwidth]{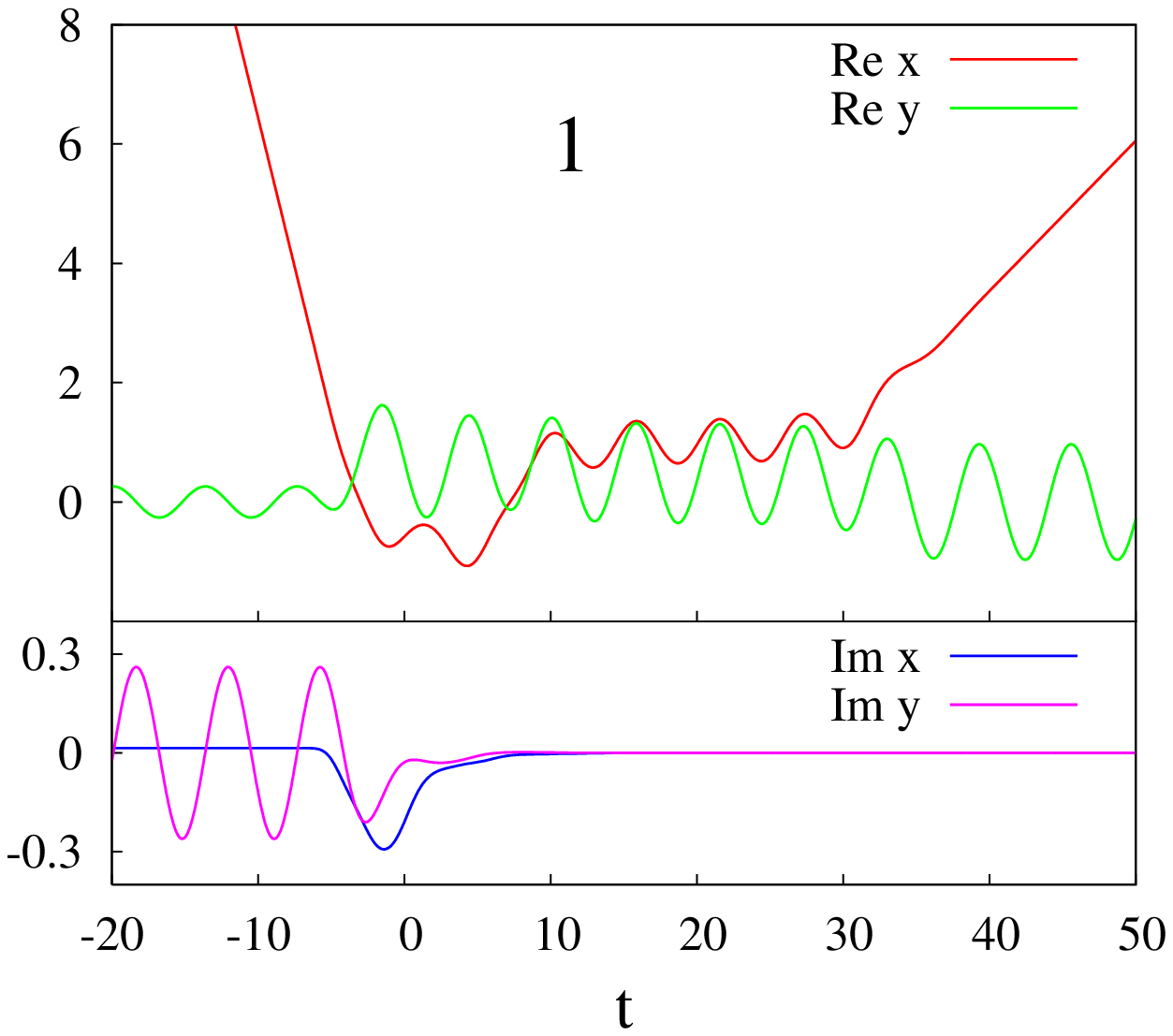}}
\centerline{\includegraphics[width=0.68\columnwidth]{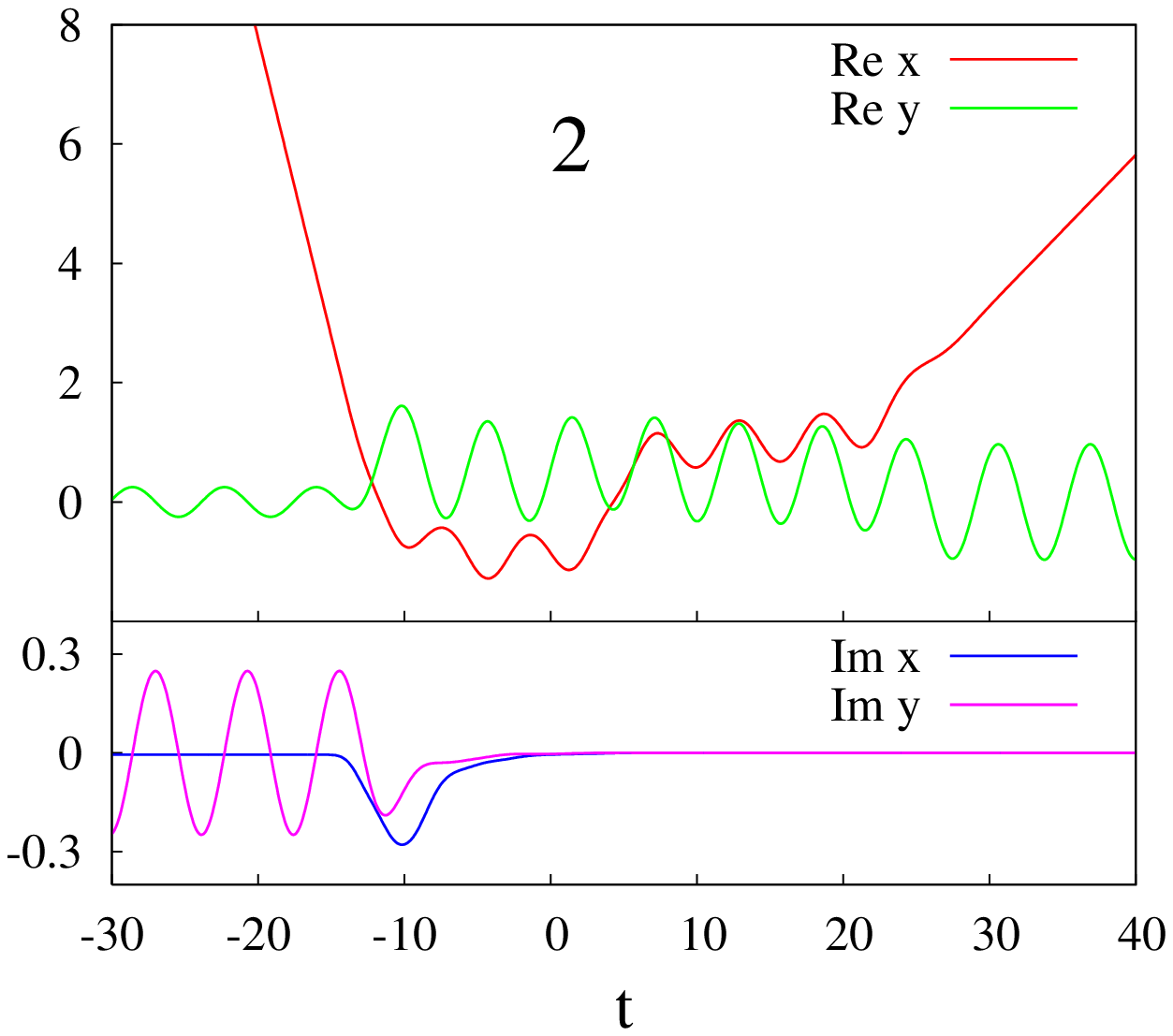}}
\centerline{\includegraphics[width=0.68\columnwidth]{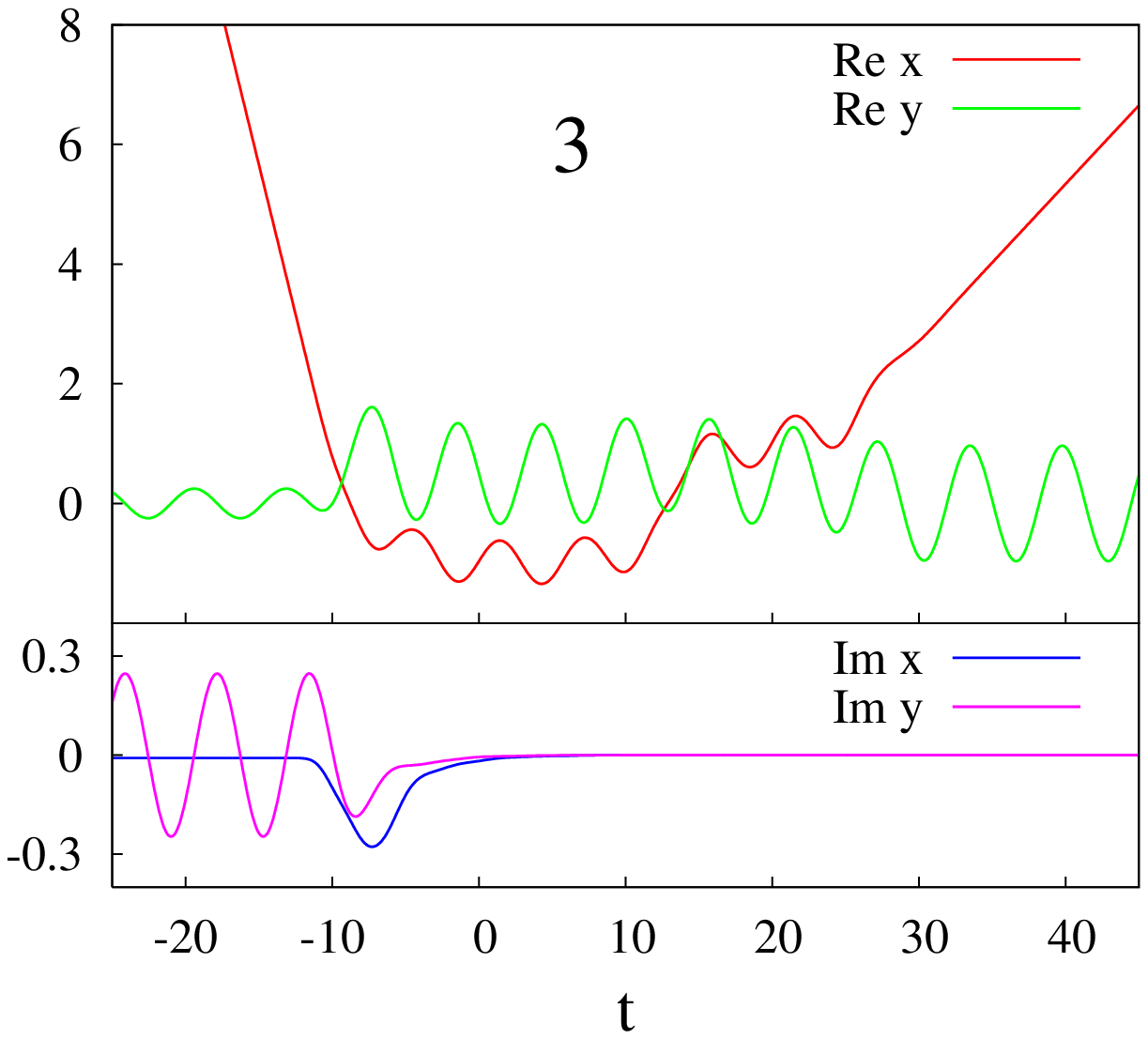}}
\centerline{\includegraphics[width=0.68\columnwidth]{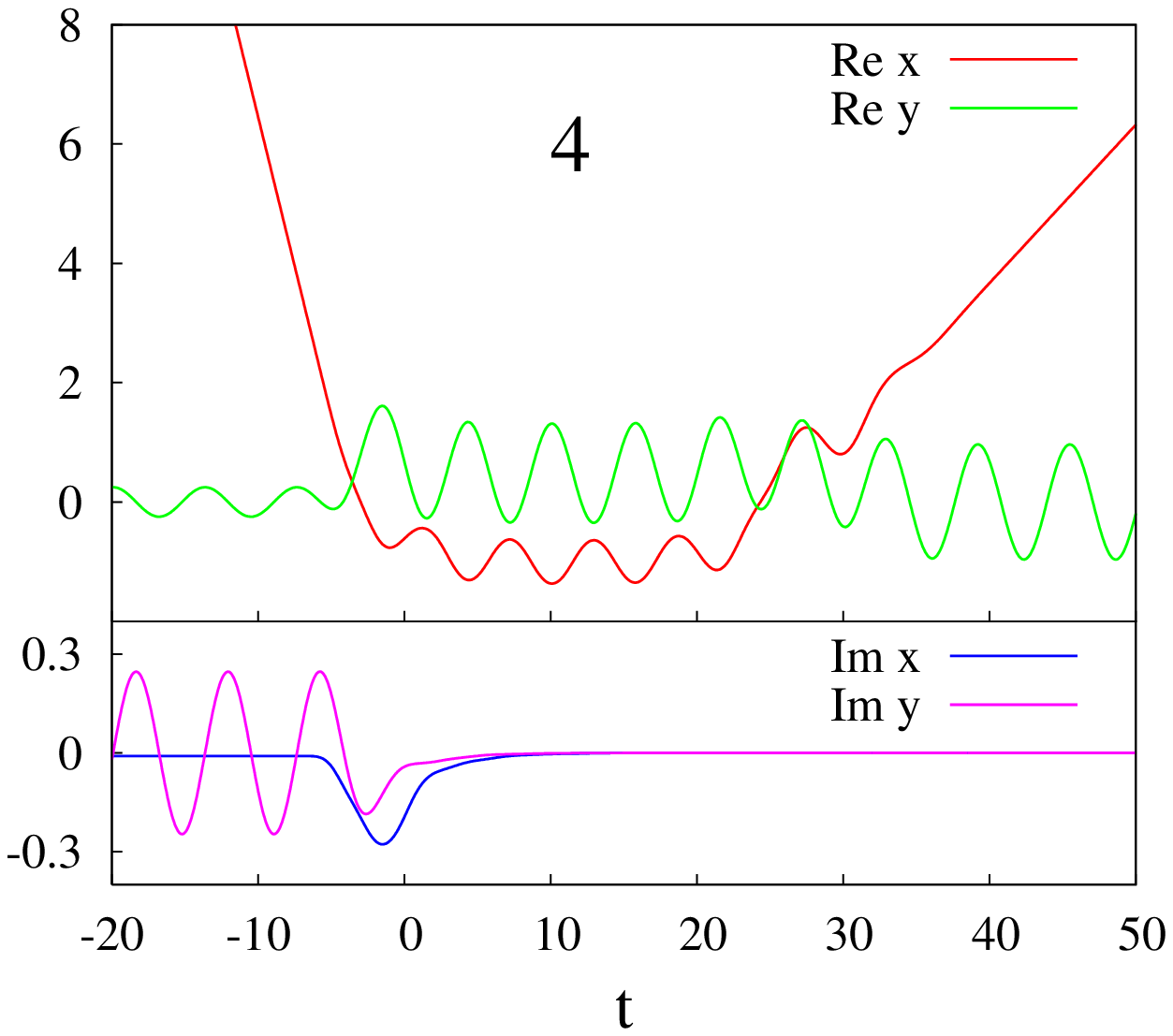}}
\caption{Four tunneling solutions from the main sequence; $E=0.5$,
  $N=0$, $\epsilon=10^{-6}$.} 
\label{fig:posled}
\end{figure}

Figure~\ref{fig:posled} shows several first representatives of the
main sequence of tunneling paths at $E=0.5$, $N=0$, $\epsilon=10^{-6}$.
Similar to their classical progenitors, the branches of
tunneling trajectories may be classified according to the number of
oscillations they perform at the far sphaleron. On the contrary,
the number of oscillations at the near sphaleron is
not an invariant of the branch: as discussed above, it grows as the
regularization 
parameter $\epsilon$ gets decreased, so that the tunneling
trajectories get stuck at the near sphaleron at $\epsilon=0$. Another
important observation about the 
plots in Fig.~\ref{fig:posled} is that the imaginary parts of the
solutions are sizeable only in the beginning of the evolution. They
fall off rapidly during the oscillations at the far 
sphaleron and become small at late times. 
This qualitative feature holds for the trajectories at other 
energies as well\footnote{On the contrary, the smallness of  
  $\mathrm{Im}\, x$ at $t\to -\infty$ observed 
  in Fig.~\ref{fig:posled} is peculiar to the trajectories at
  $E=0.5$. It implies that the corresponding values of $T$ are small,
  see Eq.~(\ref{bvp3}), so the plotted trajectories are close to the
  real-time instantons.}.  
Thus, the suppression
exponents of the trajectories are saturated during the first few
oscillations at the far sphaleron, and depend weakly on
the subsequent evolution. In addition, 
the trajectories from the main sequence almost coincide in the beginning
of the process. 
\begin{table}[t!]
\begin{center}
    \begin{tabular}{|c|c|c|c|c|c|}
      \hline
       $j\,\diagdown\,E$&0.1
       &0.3
       &0.5
       &0.7
       &0.9
       \\
      \hline
      1&0.3188
      &0.1625
      &0.1098
      &0.1398
      &0.2259
      \\
      \hline
      2&0.2586
      &0.1380
      &0.0991
      &0.1341
      &0.2221
      \\
      \hline
      3&0.2373
      &0.1340
      &0.0979
      &0.1336
      &0.2219
      \\
      \hline
      4&0.2307
      &0.1333
      &0.0978
      &0.1336
      &0.2219
      \\
      \hline 
      5&0.2285
      &0.1331
      &0.0978
      &0.1336
      &0.2219
      \\
      \hline
      $\infty$&0.2272
      &0.1331
      &0.0978
      &0.1336
      &0.2219
      \\
      \hline
    \end{tabular}
\end{center}
\caption{The suppression exponents of the complex trajectories from
  the main 
  sequence at $N=0$. The rows represent the indices $j$ of the
  trajectories, while the columns correspond to the values of energy $E$.
The last row refers to the limiting solution.}
\label{table:1}
\end{table}
This implies fast convergence of
the suppression exponents $F_j$ to the limiting value $F$ according to
the formula (\ref{Flim}). 

The above convergence is demonstrated in Table~\ref{table:1},
where the suppression exponents $F_j(E)$ are presented
for several values of energy $E$ at $N=0$. As expected from the
heuristic argument, the value of $F_j$ decreases as $j$ gets larger. 
The limiting values $F(E)$ are also included into
Table~\ref{table:1}. 
\begin{figure}[t]
\begin{center}
\includegraphics[width=0.7\columnwidth]{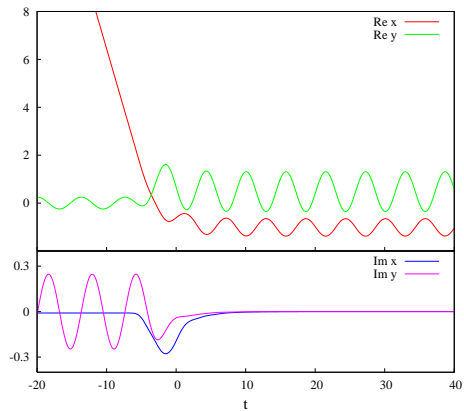}
\end{center}
\caption{The limit of the main sequence of tunneling trajectories:
the solution which gets stuck at the far
sphaleron; $E=0.5$, $N=0$.} 
\label{fig:predel}
\end{figure}
They can be obtained by
extrapolating the dependences of the suppression exponents $F_j$ on $j$,
which are well fitted by the formula
\[
F_j=F+a\e^{-bj}\;,
\]
where $a$, $b$ are real positive coefficients. A better
way, which is exploited in this paper, is to find the limit $j\to
+\infty$ of the tunneling solution itself, and then calculate the
limiting value of the suppression exponent using Eq. (\ref{suppr1}). 
The limiting solution performs an infinite number of oscillations on
the far sphaleron; it can be obtained numerically by the method 
described in App. A. The limiting solution at $E=0.5$, $N=0$ is
shown in Fig.~\ref{fig:predel}.

So far we have considered only the tunneling trajectories from the
main sequence and demonstrated that Eq.~\eqref{Flim} reproduces the lower
bound  of their suppressions. 
\begin{figure}[!ht]
\centerline{\includegraphics[width=0.68\columnwidth]{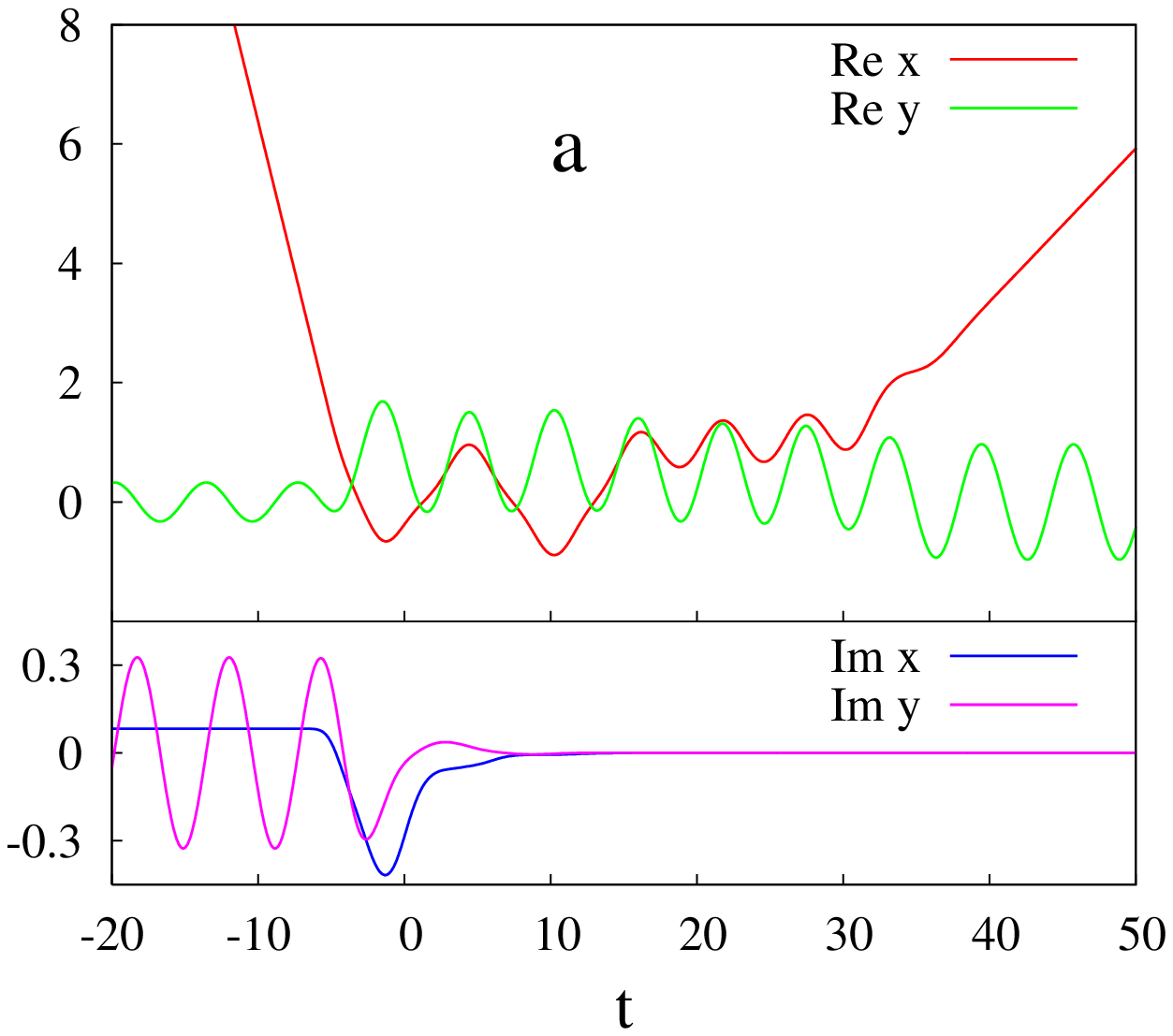}}
\centerline{\includegraphics[width=0.68\columnwidth]{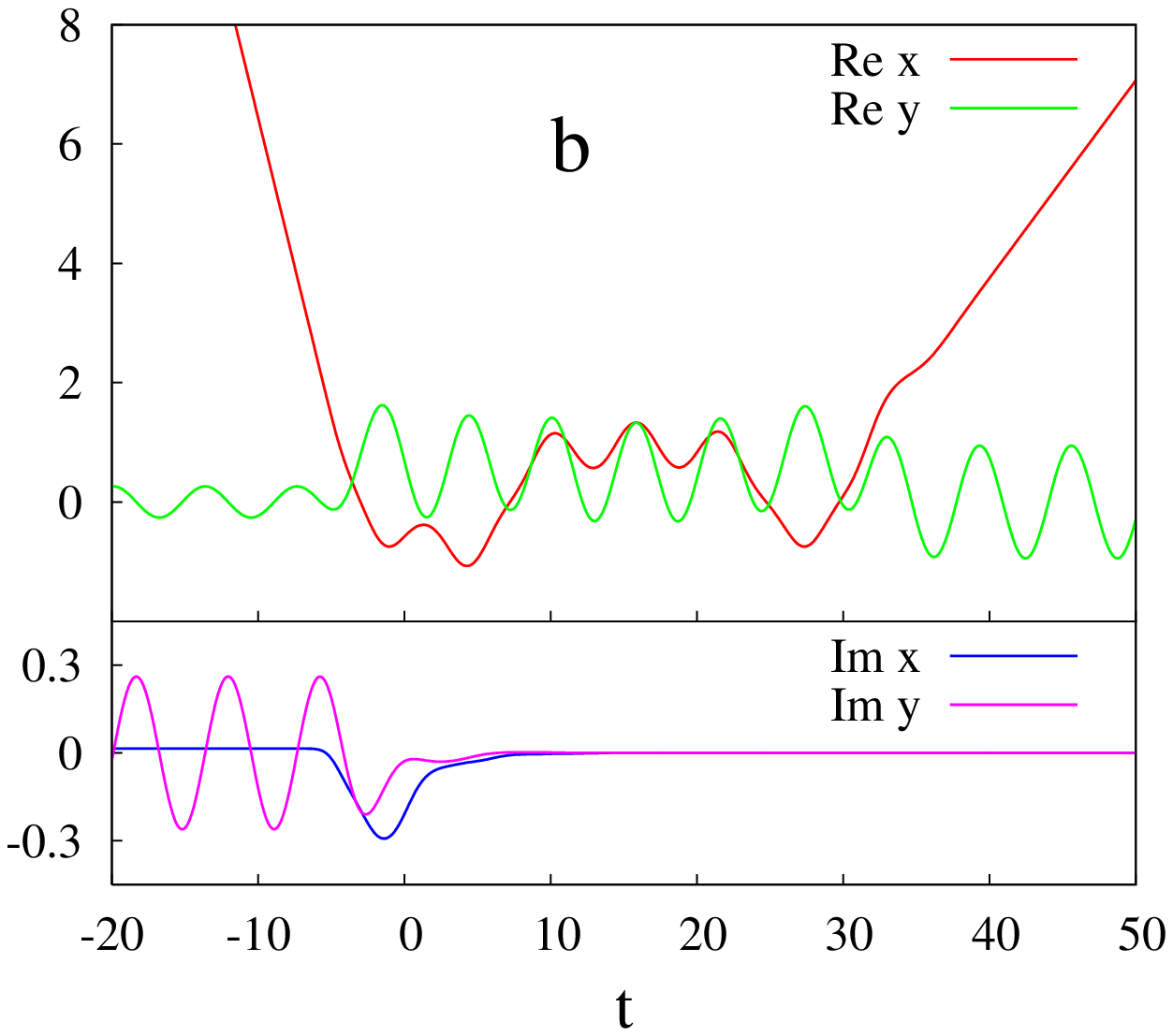}}
\centerline{\includegraphics[width=0.68\columnwidth]{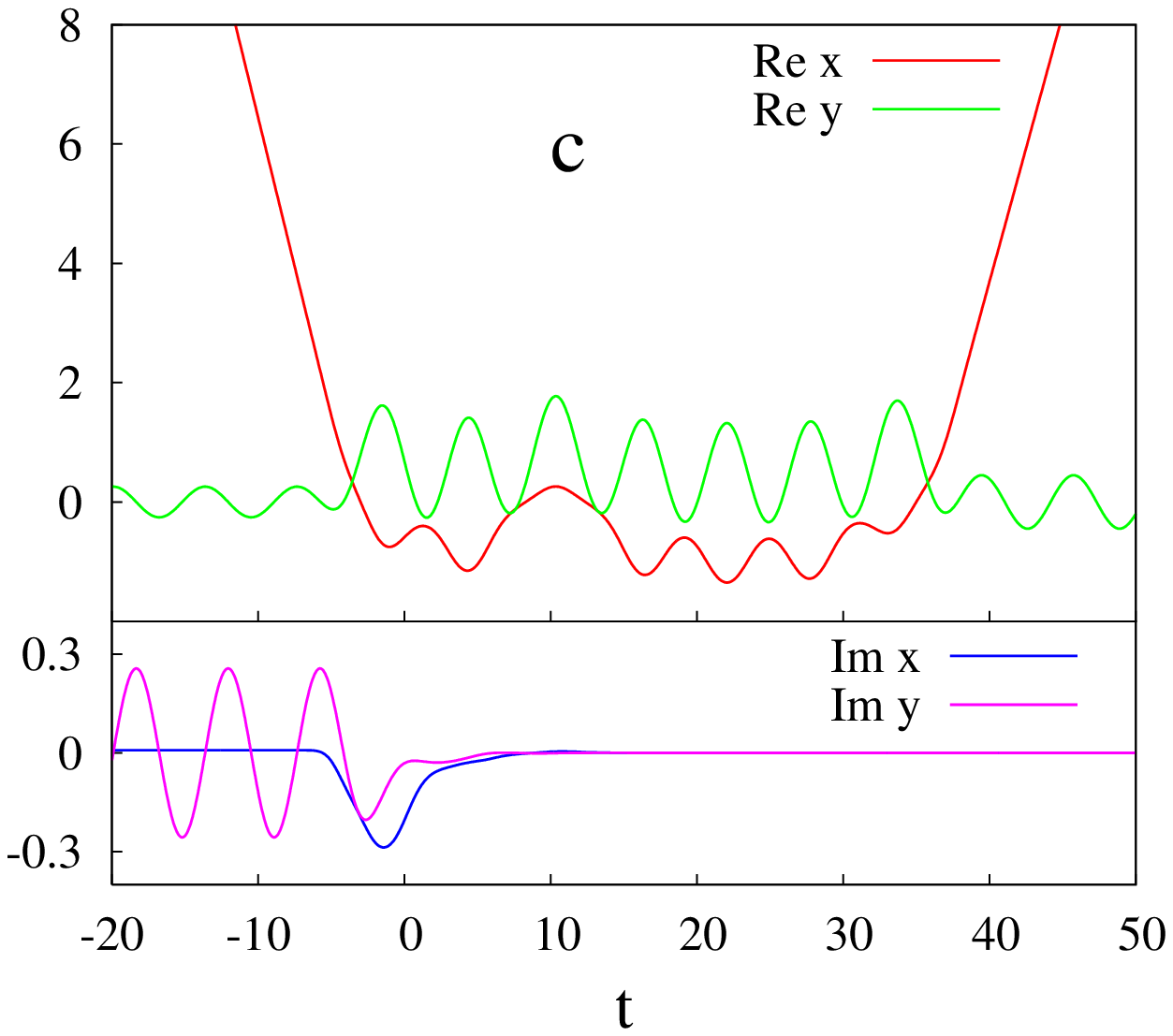}}
\centerline{\includegraphics[width=0.68\columnwidth]{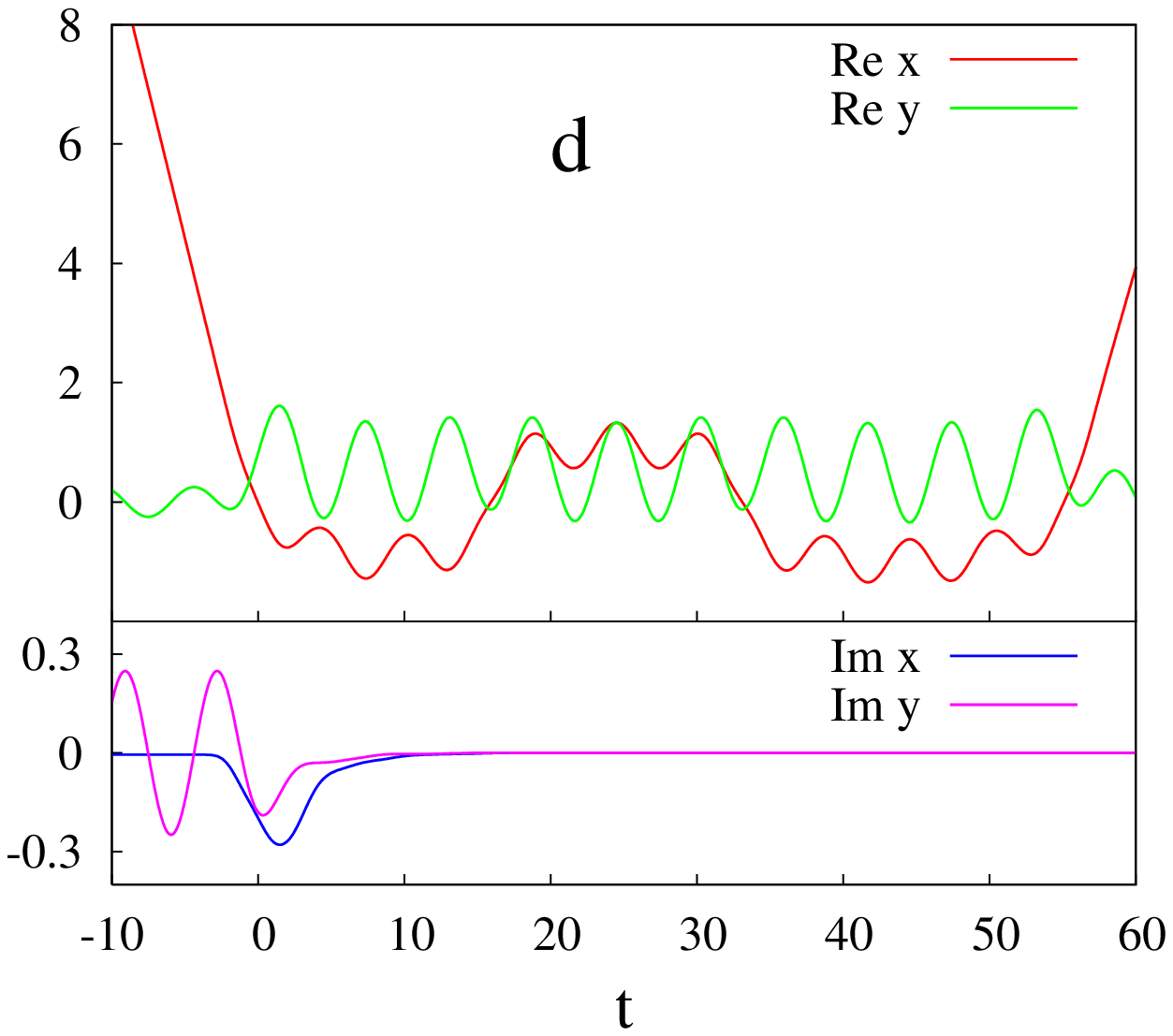}}
\caption{The examples of secondary tunneling trajectories; $E=0.5$,
  $N=0$, $\epsilon=10^{-6}$.} 
\label{fig:other}
\end{figure}
We have checked that the limiting value  $F(E)$
is lower than the suppression exponents of other tunneling
trajectories as well. As an
illustration, consider the tunneling trajectories shown in
Fig.~\ref{fig:other}. The
differences between their suppression exponents  and that of the
limiting solution are plotted in Fig.~\ref{fig:diff}. 
\begin{figure}[t]
\begin{center}
\includegraphics[width=0.95\columnwidth]{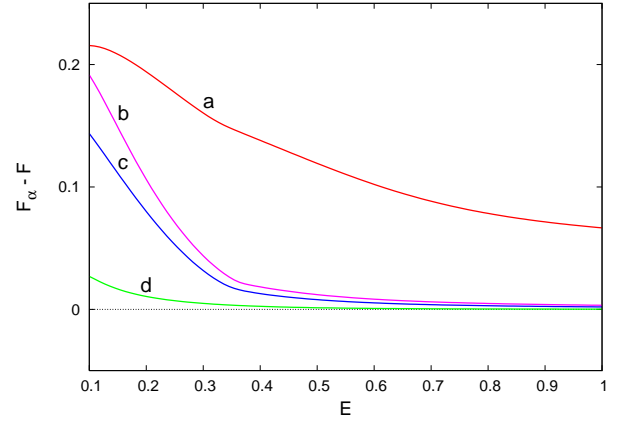}
\end{center}
\caption{The difference between the suppression exponents of the secondary
  trajectories of the types 
  shown in Fig.~\ref{fig:other}  and that of the limiting
  solution; $N=0$.} 
\label{fig:diff}
\end{figure}
The limiting solution is evidently the least suppressed one.   

Let us discuss the physical interpretation of the obtained results.
The form of the limiting solution implies that the reflection
process in our model proceeds in two stages. 
First, the far sphaleron state
gets created. This stage is exponentially suppressed due to the 
essential modification of the particle state needed for the
sphaleron creation. Second, the sphaleron state decays into
the asymptotic region $x\to+\infty$ with probability of order
$1$. This two-stage process is a manifestation of the 
phenomenon of ``tunneling on top of the
barrier'' \cite{Takahashi2003,Bezrukov:2003yf}; the phenomenon 
is generic for the inclusive
tunneling processes, see 
Refs. \cite{Bezrukov:2003er,Levkov:2004tf} for the field theoretical
examples.  

Remarkably, though we came to the limiting solution considering the
accumulation point of an infinite number of complicated tunneling
paths, the solution itself is unique and very simple. All the chaotic
features of motion, like going back and forth between the
sphalerons, are related to the second stage of the
reflection process: the decay of the far sphaleron. As the second stage is
unsuppressed, one might conclude that, after all, the chaotic motions
are irrelevant for the calculation of the main suppression
exponent. Note, however, that one is not able to guess from the
beginning, which tunneling solution corresponds to the smallest
suppression; therefore, the systematic  analysis of the 
chaotic motions is essential.  Besides, the sub-dominant contributions
of the chaotic trajectories might be important for the analysis of 
the process at finite values of the 
semiclassical
parameter $g$~\cite{Adachi:1986,Shudo:1996,Onishi2003}. 

\begin{figure*}[t]
\begin{center}
\includegraphics[width=0.7\textwidth]{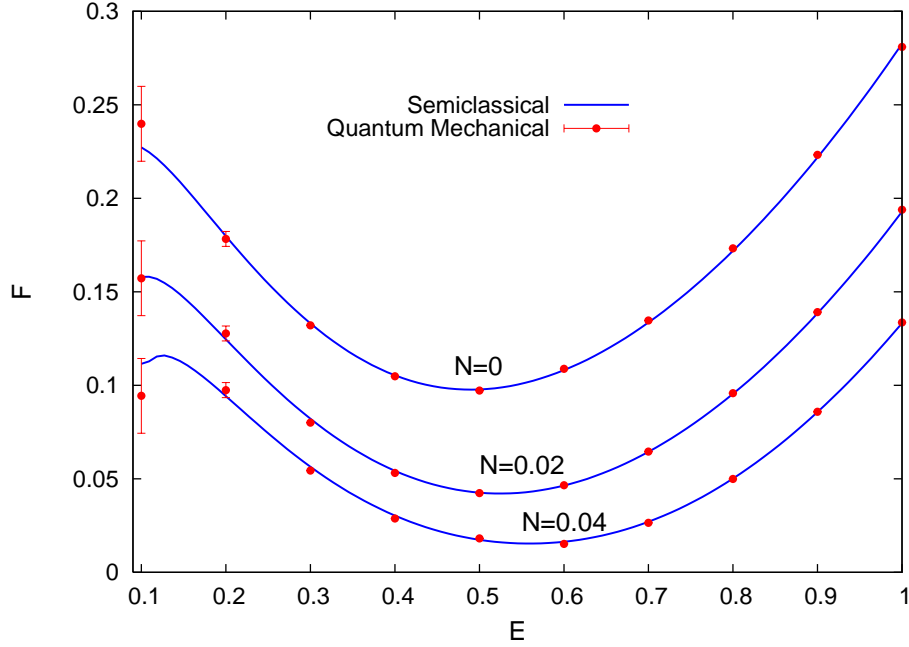}
\end{center}
\caption{The semiclassical (lines) versus quantum mechanical (points)
  results for the suppression exponent $F(E)$ at three different
  values of $N$. The errors of the quantum mechanical computations are
  smaller than the point size for $E\geq 0.3$.} 
\label{fig:main}
\end{figure*}

The final semiclassical results for the dependence
of the suppression exponent $F$ on energy $E$ for 
$N=0;\, 0.02;\, 0.04$ 
are presented in Fig.~\ref{fig:main}. 
Note that the suppression exponent 
is non-monotonic function of energy 
with the minimum near 
$E=0.5$.\footnote{
In the range $E<0.1$, which is not shown in
  Fig.~\ref{fig:main}, the suppression attains the local maximum and goes
  down to zero as $E$ decreases toward the boundary of the
  classically allowed region.} 
We remind that the minima of $F(E)$ correspond to the particularly
interesting solutions, real--time instantons. 

\section{Exact quantum computations}
\label{sec:5}

In this Section we extract the suppression exponent $F$ from the 
exact reflection probability (\ref{rcoef}). One begins by solving
numerically the Schr\"odinger equation 
\begin{equation}
\label{eq:13}
{\cal H}|\psi \rangle = {\cal E}|\psi \rangle\;.
\end{equation}
It is convenient to work in the original variables
$X$, $Y$, $P_X$, $P_Y$, see Eq. (\ref{eq:20}), in order to bring the
kinetic term of the Hamiltonian into the canonical form. One rewrites 
Eqs. (\ref{ham1}), (\ref{ax}) as
\be
\label{ham3}
{\cal H}=\frac{P_X^2+P_Y^2}{2}+\frac{1}{2}(Y-\tilde a(X))^2\;,
\ee
where
\be
\label{newa}
\tilde a(X)=\frac{1}{g}a(gX)\;.
\ee
Basically, our numerical technique follows the
lines of Refs.~\cite{Bonini:1999kj,Levkov:2002qg}. 
One works in the asymptotic basis formed by a direct product
of the translatory coordinate eigenfunctions $|X\rangle$ and
eigenfunctions $|n\rangle$ of the $y$--oscillator with the 
fixed frequency
$\omega=1$,  
$$
\psi_n(X) = \langle X,\, n| \psi\rangle\;.
$$
The stationary Schr\"odinger equation (\ref{eq:13})
reads, 
\begin{equation}
\label{eq:1}
\frac{d^2}{dX^2}\psi_n(X) = \sum_{n'} A_{nn'}(X) \psi_{n'}(X)\;,
\end{equation}
where 
\begin{multline*}
A_{nn'}(X)  = \langle n | \hat{P}_Y^2 
+ (\hat{Y} -\tilde a(X))^2 - 2{\cal E}
|n'\rangle \\
= \delta_{n,n'}(2n+1+\tilde a^2(X)-2{\cal E}) \\
- \tilde a(X)(\delta_{n,n'-1}\sqrt{2n+2}
  +  \delta_{n,n'+1}\sqrt{2n})
\end{multline*}
is an infinite three--diagonal matrix.

In the asymptotic regions $X\to \pm \infty$, $\tilde{a}(X)\to 0$ the
interaction terms become negligibly small, and the solution takes the
form, 
\begin{equation}
\label{eq:4}
\psi_n(X) \to r_n^{\pm} \mathrm{e}^{iP_nX} + t_n^{\pm}
\mathrm{e}^{-iP_nX}\;,
\end{equation}
where 
\begin{equation}
\label{eq:5}
  P_n = \sqrt{2{\cal E} - (2n + 1)}
\end{equation}
stands for the asymptotic translatory momentum of the $n$-th mode. 
The boundary conditions for the stationary wave function $\psi_n(X)$ are
constructed in the standard way. The particle comes from the
right, $X\to +\infty$, in the ${\cal N}$--th oscillator state; hence, we fix
\begin{equation}
\label{eq:2}
t_n^{+} = \delta_{n,{\cal N}}\;.
\end{equation}
On the other hand, only the outgoing wave should remain at 
$X\to -\infty$,
\begin{equation}
\label{eq:3}
r_n^{-} = 0\;.
\end{equation}
Note that while the low--lying modes are oscillatory at the 
asymptotic, the
ones with $n > {\cal E}-1/2$ grow (decay) exponentially, see
Eq. (\ref{eq:5}). Physically, the latter correspond to the
kinetically 
inaccessible region, where the energy of the transverse oscillations
exceeds ${\cal E}$. We fix the boundary conditions for 
them by killing the parts growing exponentially toward
infinities. One notes that after the proper continuation of 
Eq. (\ref{eq:5}), 
\begin{equation}
\label{eq:6}
P_n = i \sqrt{(2n+1) - 2{\cal E}}\;, \qquad n > {\cal E}-1/2\;,
\end{equation}
the aforementioned conditions coincide with Eqs. (\ref{eq:2}),
(\ref{eq:3}).  

Equations (\ref{eq:1}), (\ref{eq:2}), (\ref{eq:3}) constitute the
boundary value problem to be solved numerically. After the stationary
wave function $\psi_n(X)$ is found, one calculates the probability
current  
\begin{equation}
\label{eq:7}
J = \mathrm{Im}\,\sum_n  \psi_n^*(X) \frac{d}{dX} \psi_n(X)\;,
\end{equation}
and hence the reflection probability
\begin{equation}
\label{eq:8}
{\cal P} = \frac{|J^{(\mathrm{out})}|}
{|J^{(\mathrm{in})}|} = \sum_{n< {\cal E}-1/2} \frac{P_n}{P_{\cal N}} 
|r_n^+|^2\;,
\end{equation}
where the currents $J^{(\mathrm{in})}$ and 
$J^{(\mathrm{out})}$ are computed with the incoming and outgoing
  parts of the wave function at $X\to +\infty$, respectively. 

The details of the numerical formulation of the problem (\ref{eq:1}),
(\ref{eq:2}), (\ref{eq:3}) are presented in 
App.~\ref{app:B}. Let us discuss the results.
\begin{figure}[htb]
\centerline{\includegraphics[width=\columnwidth]{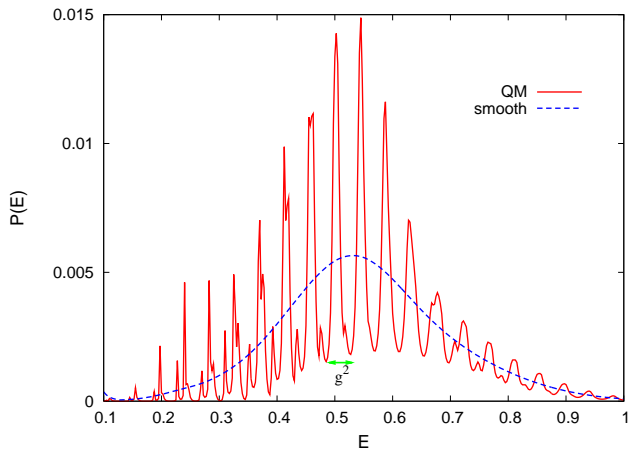}}
\caption{The function ${\cal P}(E)$
  plotted for $N=0$, $g =0.2$. 
  The dashed curve represents the smoothened result, 
  Eq.~(\ref{eq:15}).
  }
\label{fig:4}
\end{figure}
The typical dependence of the reflection probability ${\cal P}$ on energy
$E=g^2{\cal E}$ is shown in Fig.~\ref{fig:4}, solid line (the values of
the other parameters are $N \equiv g^2{\cal N} = 0$, $g = 0.2$). 
The striking feature is that the graph is modulated by sharp
oscillations. At first
glance, this picture is incompatible with the results of the
semiclassical analysis, where we obtained, at least in the leading
order, that the tunneling probability, 
${\cal P}\propto \e^{-F(E)/g^2}$, is a smooth function of 
energy. We are going to demonstrate the opposite: the quantum mechanical
results reconcile nicely with the semiclassical ones.

By computing the reflection probability at different $g$, one notices
the following important properties. First, 
the period of oscillations scales like $\tau_E\propto g^2$. So, in
the semiclassical limit $g\to 0$ the oscillations become more and more
frequent. Second, one considers $F_g \equiv g^2 \ln {\cal P}$ and
asks whether the  amplitude of oscillations of this quantity goes to
zero in the limit $g\to 0$. We 
certainly observed that it drops down as
$g$ decreases though we were unable to figure out\footnote{This is due
  to the limitations of the numerical approach: one cannot obtain
  solutions to the Schr\"odinger equation at arbitrarily small $g$,
  see App.~\ref{app:B}.} whether it indeed vanishes at  
$g\to 0$. The properties mentioned above suggest the
interpretation of the oscillations as the result of the quantum
interference between the contributions of various tunneling
trajectories found in Sec.~\ref{sec:4}. Suppose for simplicity
that there are only two such trajectories. Let us denote their complex
actions by $S_1(E)$ and $S_2(E)$ (we omit  
the dependence on $N$).
Then, the total
reflection amplitude reads,  
$$
r \sim A_1 \mathrm{e}^{iS_1(E)/g^2} + 
A_2\mathrm{e}^{iS_2(E)/g^2}\;.
$$
For the sake of argument  we suppress the index $n$ and disregard
the initial--state contributions into the exponents. In the above formula
$A_1$ and $A_2$ 
stand for the pre-exponential factors of the partial processes. For
the reflection probability one writes,
\begin{multline}
\label{eq:14}
{\cal P} \propto |r|^2 \sim |A_1|^2 
\mathrm{e}^{-2\mathrm{Im}\,S_1(E)/g^2} + 
|A_2|^2 \mathrm{e}^{-2\mathrm{Im}\,S_2(E)/g^2} \\
+2|A_1 A_2|
\mathrm{e}^{-\mathrm{Im}\,(S_1(E)+S_2(E))/g^2}\times\\
\times \cos\left[\mathrm{Re}\, (S_1(E) - S_2(E))/g^2 + 
\mathrm{arg}(A_1/A_2)\right] \;.
\end{multline}
Along with the terms corresponding to the probabilities of
the partial processes, one gets the interference term, which
results in oscillations with period $\tau_E$ 
of order $g^2$. Of course,
if one of the solutions, say, the first one, is dominant,
$\mathrm{Im}\, S_1 < \mathrm{Im}\, S_2$, the relative contributions of
the two last terms in
Eq. (\ref{eq:14}) vanish exponentially fast at $g\to 0$. 
In our case the situation is more subtle, however. As the
semiclassical analysis reveals, in the model under consideration there
exists an infinite number of tunneling paths,
which pile up near the dominant 
limiting solution. So, at each finite value of $g$ there are solutions
which satisfy
$\mathrm{Im}\,(S_2 - S_1) \lesssim g^2$
(the index ``$1$'' still marks the dominant solution here). Thus,
the sum (\ref{eq:14}) always contains an infinite number
of oscillating terms
producing a complicated interference pattern; it is not
clear whether the oscillations disappear at small $g$.

What saves the day is the aforementioned scaling of the
oscillation period. Indeed, $\tau_E$ vanishes at $g\to 0$
implying that the oscillations become indiscernible in 
the semiclassical limit. One obtains a quantity which is well-behaved
in this limit by averaging the reflection probability over 
several periods of oscillations. To be more precise, we consider the
smoothened probability 
\begin{equation}
\label{eq:15}
{\cal P}^{(s)}(E,N) = \int d E' \, {\cal
  D}_{\sigma} (E - E') {\cal P}(E',N)\;,
\end{equation}
where ${\cal D}_\sigma$ is the bell-shaped function,
$$
D_\sigma(E) = \frac{\mathrm{e}^{-E^2/\sigma^2}}
{\sigma\sqrt{\pi}} \;, \qquad \int d
E\, D_\sigma(E) = 1\;.
$$
If $\sigma  = g^2\Sigma$, where $\Sigma$ is a fixed number, the
smoothening (\ref{eq:15}) does not spoil the value of the dominant
suppression exponent. Indeed, for the first term in Eq. (\ref{eq:14})
one writes,
\begin{align}
\label{eq:17}
{\cal P}^{(s,\mathrm{dom})} &\propto \int dE'\, 
|A_1|^2\mathrm{e}^{-F_1(E')/g^2}{\cal D}_\sigma(E'-E) \\ \notag
&\approx |A_1|^2
\mathrm{e}^{{F_1'}^2(E)\Sigma^2/4}\cdot 
\mathrm{e}^{-F_1(E)/g^2}\;,
\end{align}
where\footnote{For the sake of argument we, again, disregard
  the boundary terms in the suppression exponent,
  c.f. Eqs.~(\ref{suppr1}).} 
$F_1 = 2\mathrm{Im}S_1$, $F_1' = dF_1/dE$, and we made
use of the Taylor series expansion in the second equality. 
It is clear that only the pre-exponential factor is
affected by the integration (\ref{eq:15}), while the exponent
$F_1(E)$ is left intact.
On the other hand, at large enough 
values of the coefficient $\Sigma$ the formula (\ref{eq:15})
represents averaging over many oscillatory periods, which 
kills all the oscillating contributions. Indeed, consider the
typical interference term,
\begin{multline}
\label{eq:16}
{\cal P}^{(\mathrm{osc})} \propto C(E) \mathrm{e}^{-F_0(E)/g^2}\times\\\times
 \cos\left[\Delta S(E)/g^2 -
  \Delta\phi(E)\right]\;,
\end{multline}
where $F_0 = \mathrm{Im}(S_1 + S_2)$, $\Delta S = \mathrm{Re}(S_1 -
S_2)$, $C = 2|A_1A_2|$, $\Delta \phi = \mathrm{arg}(A_1/A_2)$. 
Performing integration, one obtains,
\begin{multline*}
{\cal P}^{(s,\mathrm{osc})}(E) \propto
C \mathrm{e}^{-F_0/g^2} \cos \left[\Delta
  S / g^2 + \Delta \phi - F_0' \Delta
  S' \Sigma^2/2\right] \\ \times\exp\left\{
\frac{\Sigma^2}{4}(F_0'^2 - \Delta S'^2)\right\}\;.
\end{multline*}
Taking into account that $F_0'\approx F_1'$ we see that 
the
interference term is suppressed by the additional factor $\exp\{-\Sigma^2
 \Delta S'^2/4\}$ with respect to the dominant contribution
 (\ref{eq:17}). 
Below we fix $\Sigma=1$. We observed that
the
interference patterns get multiplied by $10^{-3}$ in this case; the latter
number is accepted as the precision of the smoothening. The graph of
the function ${\cal P}^{(s)}(E)$ is shown in Fig.~\ref{fig:4}, dashed
line.

Our final remark concerns the physical meaning of the
smoothening procedure. In a realistic experiment one cannot
fix the energy ${\cal E}$ of the incoming particles exactly; rather, 
one works
with some sharply peaked energy distribution ${\cal D}$ of a width
$\Delta {\cal E} \sim \Sigma$. Formula (\ref{eq:15}) represents 
averaging over such distribution. 

Now, we are ready to consider
the limit $g^2\to 0$. Our aim is to check the following asymptotic
formula, cf. Eq.~(\ref{probab}),
\begin{equation}
\label{eq:18}
{\cal P}^{(s)}(E,N) \to g^\gamma A(E,N) \mathrm{e}^{-F(E,N)/g^2} 
\;\;\mathrm{as} \;\; g^2\to 0\;.
\end{equation}
We compute the value of the quantity $-g^2\ln
{\cal P}^{(s)}$ at several  $g^2\ll 1$, keeping $E$
and $N$ fixed, and fit the graph with the expression
\begin{equation}
\label{eq:19}
-g^2 \ln{{\cal P}^{(s)}} = F -\gamma g^2 \ln g - g^2 \ln{A}\;.
\end{equation}
\begin{figure}[htb]
\centerline{\includegraphics[width=0.9\columnwidth]{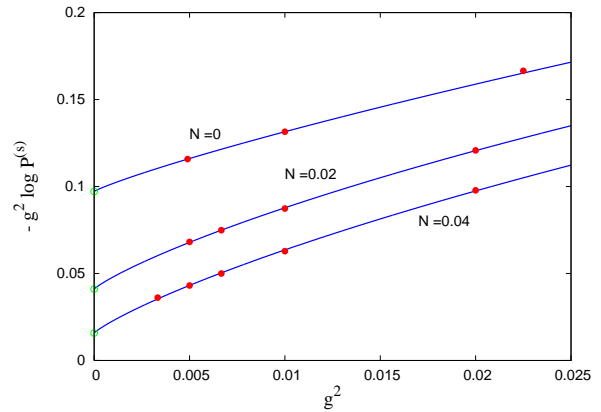}}
\caption{The quantity $-g^2\ln {\cal P}^{(s)}$ viewed as a function of
  $g^2$ at fixed $N$ and $E = 0.5$. } 
\label{fig:5}
\end{figure}
It is important to point out that we do not consider $\gamma$ as a
free parameter of the fit. Rather, we use the 
following result of Ref.~\cite{prefactor}:
for the 
sphaleron--mediated processes, such as ours, $\gamma=1$ at
$N=0$ (vacuum initial state), and $\gamma=2$ at $N\ne 0$. 
Some of the numerical results (points) together with their fits by
the formula (\ref{eq:19}) (lines) are shown in
Fig.~\ref{fig:5}. The graphs are drawn for three different values of
$N$ and $E=0.5$. Apparently, our data are well
approximated by the asymptotic (\ref{eq:19}). To get the value of the
suppression exponent $F$, one extrapolates the curves in
Fig.~\ref{fig:5} to the point $g^2=0$, where the quantum mechanical
results should coincide with the semiclassical ones.  
The error of the extrapolation arises mainly from the
disregarded terms proportional to the higher powers of $g^2$. 
One reduces such errors by computing the reflection
probability at the smallest possible $g^2$. In practice we used two values
of the semiclassical parameter in each fit, namely, $g = 0.1,\; 0.07$
at $N=0$, and $g=0.1,\; 0.08$ at $N=0.02,\; 
0.04$. The extrapolation error is determined by pouring some
additional points into the fit; it varies from $\delta F \sim 10^{-2}$
at small $E$ to $\delta F \sim 10^{-4}$ at $E\sim 1$. It matches
with the estimate $O(g^4)\sim 10^{-4}$  of higher--order terms of the 
semiclassical expansion which are neglected in Eq. (\ref{eq:19}).

Our final results for the suppression exponent $F(E,N)$ 
extracted from the numerical solution of the Schr\"odinger equation are
presented in 
Fig.~\ref{fig:main} (points with error bars standing for the accuracy
of the extrapolation). 
The quantum mechanical results are in  very good 
agreement with the semiclassical ones (lines). This
justifies the semiclassical approach presented in this paper.

\section{Summary and discussion}
\label{sec:6} 
In this paper we tested the semiclassical method of complex
trajectories in the regime of chaotic 
dynamical tunneling. We studied a
particular example,  
over-barrier reflection in the two-dimensional 
waveguide model
(\ref{ham1}). 
The initial state of the process was fixed by the total energy 
${\cal E}$ and occupation number ${\cal N}$ of the transverse
oscillatory motion.
We calculated the suppression exponent of the process both 
semiclassically and by solving exactly the full Schr\"{o}dinger
equation. The two approaches show very good agreement. 

The tunneling trajectories 
in our semiclassical approach are obtained as solutions 
to the boundary value
problem
(\ref{bvp}). We advocated a particular method for finding these
solutions. It consists of two important ingredients:\\
(i) determination of the solution at some values of the
initial--state parameters $E$, $N$;\\ 
(ii) gradual deformation of the solution to other $E$, $N$.

The deformation procedure
(ii) was developed in
Refs. \cite{Kuznetsov:1997az,Bonini:1999kj}. The 
advantages of this procedure are the simplicity of numerical
implementation and generality. It can be applied efficiently to 
systems with many degrees of freedom,
including non--trivial models of field theory, see
Ref.~\cite{Bezrukov:2003er}. 

On the other hand, a generic approach for performing the step (i) was
missing so far. In this paper we proposed the systematic procedure which
fills this gap. The procedure enables one to obtain
the complex tunneling trajectories
starting from the real classical solutions. It is based on the
$\epsilon$--regularization method of
Refs. \cite{Bezrukov:2003yf}.
Our procedure appears to be generic.
It can be applied to 
any process which proceeds classically at some values of
the initial--state parameters (at large $N$ in our case) and becomes
exponentially suppressed at other values. 

The process we studied is a particular example of chaotic tunneling.
The chaoticity manifests itself in the infinite number of tunneling
solutions. Our procedure, which connects the tunneling solutions to the
classical ones, turns to be highly efficient in this situation. Namely, it
enables one to classify the tunneling trajectories on the basis of the
analysis of the classical dynamics. This was demonstrated explicitly in
the present paper. 

In addition, we proposed a heuristic criterion for
sorting out the least suppressed tunneling trajectories basing 
on the
classification of their classical progenitors. Hopefully, this
criterion will be useful for the processes in other dynamical systems
as well. 

Another interesting feature of our setup is the phenomenon of optimal
tunneling. Namely, the suppression exponent $F$  considered as a function of
energy $E$ at fixed $N$ is non-monotonic.
It attains the local minimum at $E\approx 0.5$ which is thus
(locally) the optimal energy for tunneling. It is worth stressing that
this behavior of the suppression exponent is unrelated to the quantum
interference, which was neglected in the semiclassical analysis and
eliminated from the exact quantum computations. Another example of
tunneling process with non-monotonic dependence of the suppression
exponent  on energy is considered in  Ref.~\cite{wigles}. 

Let us mention some open issues. 

The method of complex
trajectories is known to suffer, in general, from the 
difficulties related to the Stokes
phenomenon~\cite{Berry:1972,Adachi:1986}. The essence of this
phenomenon is that solutions of the tunneling problem may be
unphysical in some regions of the parameter space;  the contributions
of such solutions to the tunneling probability should be dropped
out. Presently, there are no  generic criteria for distinguishing
between the physical and unphysical solutions. One notes 
that the method we put forward in this paper trivially excludes some
of the unphysical solutions. Indeed, we start from the classically
allowed region of initial data (large $N$) where the  physical
solutions are precisely the real--valued classical trajectories. At
the second step we relate these  solutions to the complex tunneling
trajectories; thus, we exclude the branches of unphysical solutions
which are complex--valued deep inside the classically allowed region of
initial data. Due to this ``automatic'' criterion, we did not see any
manifestations of  
the Stokes phenomenon in the semiclassical calculations 
presented above. We remark, however, that the tunneling process we studied
is the simplest one from the point of view of the Stokes phenomenon;  the
above ``automatic''exclusion of unphysical solutions is
insufficient in somewhat more involved situations. Two
aspects of our model simplify the analysis. First, we
observed that the tunneling solutions obtained from the real
classical trajectories of a given topology $\alpha$ form a {\it
  single} smooth branch which covers the entire range of initial 
data. Second, we showed that the hierarchy of the
suppression exponents $F_{\alpha}$ is the same at different values of
$E$, $N$. These two observations allowed us to identify a single
smooth branch of physical solutions which give the dominant
contribution to the tunneling probability. In other models (see, e.g.,
Ref.~\cite{wigles}) 
the solutions obtained by small deformations from different parts of
the classically allowed region of initial  data may correspond to
{\it different} smooth branches of complex trajectories. Each of these
branches may be physical and dominant in some region of the initial data
plane, and unphysical or sub-dominant in other regions. If this is the
case, some method of treating the Stokes
phenomenon \cite{Shudo:1996,Ribeiro:2004,Parisio:2005}
should be exploited. 

The approach adopted in this paper
was to perform the evaluation of the tunneling probability as
the systematic semiclassical expansion in terms of $g^2$. We have
calculated the leading term (suppression
exponent). It is of interest  to develop a method
for calculating the sub-leading terms, in particular, the
pre-exponential factor $A$, see Eq. (\ref{probab}). Presumably, this can
be done along the lines of Ref.~\cite{prefactor}. An important problem
which should be solved here is how to cope with the
infinite number of  tunneling paths, whose suppressions can be
arbitrarily close to the limiting value. In particular, one has to
understand whether the contributions of all the paths should be summed
up or the correct value of the prefactor is determined by a sort of
limiting procedure similar to the one we used to obtain the suppression
exponent. 

Another question is the following one. 
While solving the Schr\"{o}dinger
equation we observed that at finite $g^2$ 
the dependence of the exact tunneling
probability on energy is modulated by
oscillations.  We conjectured that they result 
from the quantum interference of different tunneling trajectories. 
In the true semiclassical limit $g^2\to 0$ the above oscillations become
infinitely frequent and should be averaged over. In real physical
situations one deals, however, with small but finite values of the
semiclassical parameter $g^2$. 
Thus it would be interesting to reproduce the interference pattern
mentioned above in the semiclassical approach. 
As suggested by 
Refs.~\cite{Shudo:1996,Shudo1995_1,Onishi2003} 
this could be done by summing up  
the contributions of various tunneling
trajectories at finite $g^2$. We leave this
investigation for the future work.

\paragraph*{Acknowledgments} We are indebted to F.L.~Bezrukov and
V.A.~Rubakov for useful discussions and helpful suggestions. This work
was supported in part 
by the RFBR grant 05-02-17363, Grants of the President of Russian
Federtion NS-7293.2006.2 (government contract 02.445.11.7370), 
MK-2563.2006.2 (D.L.), MK-2205.2005.2 (S.S.), the Grants of the Russian
Science Support Foundation (D.L. and S.S.), the 
Fellowship of the ``Dynasty'' Foundation (awarded by the Scientific
board of ICPFM) (A.P.) and the INTAS grant YS 03-55-2362 (D.L.). 
D.L. is
grateful to the Universite Libre de Bruxelles and to EPFL, Lausanne,
for the hospitality during his visits. The numerical
calculations were performed on the Computational cluster of the
Theoretical division of INR RAS.

\appendix

\section{A method to obtain the unstable solutions}
\label{app:A}
In this Appendix we describe the numerical method used 
to obtain the limiting tunneling trajectories considered in
Sec.~\ref{sec:4}, namely, the ones ending up at
the far sphaleron at late times. The main problem here is the instability
of the trajectories in question. 

Our main idea is to
add the following term to the action functional,
\be
\label{Mterm}
S[{\boldsymbol{x}}]\mapsto  S[{\boldsymbol{x}}]+iM(x(t_f)-x_f^0)^2\;,
\ee
where $x(t_f)$ stands for the final value of the coordinate $x$, while
$M>0$, $x_f^0$ are real parameters. For a given trajectory 
the term~(\ref{Mterm}) leads to additional contribution to the 
suppression exponent,
\be
\label{MF}
\Delta F =2M(x(t_f)-x_f^0)^2\;.
\ee
The introduction of the above term induces the following
modification of the boundary conditions (\ref{bvp2}),
\bseq
\label{Mboundary}
\begin{align}
\label{Mboundary1}
&\Im x(t_f)=0\;,\\
\label{Mboundary2}
&\Im\dot x(t_f)=-2M(x(t_f)-x_f^0)\;.
\end{align}
\eseq
This can be shown using the systematic approach of
Refs. \cite{Bezrukov:2003yf,prefactor,wigles}. 
At large positive $M$ the term (\ref{MF}) fixes the value of the final
$x$-coordinate of the solutions to be close to $x_f^0$.
For our purposes 
we choose $x_f^0$ to be in the vicinity of
the far sphaleron, $x_f^0=-1$.
Solutions which approach the far sphaleron at late
times are obtained in the following way. One finds a solution to the
original equations of motion which spends finite time on
the far sphaleron,
and cuts it at the moment $t = t_f$ when it is still at $x\approx
-1$. Using this trajectory as the zeroth--order approximation, one
applies the Newton--Raphson algorithm and obtains the 
solution satisfying the boundary conditions (\ref{Mboundary}). The
latter solution is defined inside the interval $t\in
[t_i,t_f]$. The next step is to continue the solution to a
larger time  interval by gradually increasing the value of $t_f$ and
deforming the tunneling solution. As a result, one obtains the
solution living at the far sphaleron for an arbitrarily
long time.

The solution obtained in this way does not, strictly speaking, satisfy
the boundary conditions (\ref{bvp2}) of the tunneling problem. In
order to restore the original boundary conditions (\ref{bvp2}) one
should, in principle, investigate the 
dependence $x_f(x_f^0)$ and find the value of $x_f^0$ which solves the
equations $x_f(x_f^0)=x_f^0$. However, it is not necessary in our
case: at late times the limiting solution is almost real, so one obtains
$\Im \dot x_f\approx 0$ automatically.

\section{Numerical solution of the Schr\"{o}dinger equation}
\label{app:B}
Here we present the numerical formulation of the problem (\ref{eq:1}),
(\ref{eq:2}), (\ref{eq:3}). First of all, the range of the
translatory coordinate should be bounded, $-L \leq X \leq L$, 
as well as the oscillator excitation number, $n < N_y$. 
Besides, we introduce the uniform lattice with spacing
$\Delta$, 
$$
X_k = -L + (k+1)\Delta\;, \qquad k = -1,\dots,N_x+1\;,
$$
where  $N_x = -2 + 2L/\Delta$.
The Taylor series expansion gives,
\begin{multline}
\label{eq:9}
\frac{1}{\Delta^2}\left[ 
\psi_{k+1} -2\psi_k+ \psi_{k-1}
\right] =\\= \psi_k'' + \frac{\Delta^2}{12} \psi_k^{(IV)} +
O(\Delta^4)\;.
\end{multline}
By  $\psi_k''$ and $\psi_k^{(IV)}$ we denote the second and 
fourth derivatives of $\psi$ at $X=X_k$. Note that 
the index $n$ is suppressed: hereafter we use the matrix notations. From
Eq. (\ref{eq:1}) $\psi_k'' = A_k \psi_k$; for the fourth
derivative one writes,
\begin{align*}
\psi_k^{(IV)} &= \frac{1}{\Delta^2} \left[ 
\psi_{k+1}''  - 2\psi_k''+ \psi_{k-1}''
\right] + O(\Delta^2) \\&= \frac{1}{\Delta^2} \left[
A_{k+1} \psi_{k+1}- 2A_k\psi_k + A_{k-1}\psi_{k-1} 
\right] + O(\Delta^2)\;.
\end{align*}
Substituting the above expressions into Eq. (\ref{eq:9}), one gets the
forth--order Numerov--Cowling approximation for Eq. (\ref{eq:1}),
\begin{multline}
\label{eq:10}
\left(1 - \frac{\Delta^2}{12}A_{k+1}\right) \psi_{k+1}-
\left(2 + \frac{5\Delta^2}{6}A_k \right) \psi_k+\\
+\left(1 - \frac{\Delta^2}{12}A_{k-1}\right) \psi_{k-1}  = 0\;, \;\;\;\; k =
0,\dots, N_x\;.
\end{multline}
It is worth noting that Eq. (\ref{eq:10}) supports the discrete
probability current, which is conserved exactly,
\begin{equation}
\label{eq:11}
J^{(\mathrm{d})} = \frac{1}{\Delta} \mathrm{Im}\, 
\psi_k^+ \left(1-\frac{\Delta^2}{12}A_k\right)\left(1 -
\frac{\Delta^2}{12} A_{k+1}\right) \psi_{k+1}\;.
\end{equation}
This conservation law was used to estimate the round--off errors.

The boundary conditions (\ref{eq:2}), (\ref{eq:3}) are imposed at the
very last and first sites, $k=N_x+1$ and $-1$ respectively. One
notes that the asymptotic formula (\ref{eq:4}) holds in the discrete case as
well, provided the continuum dispersion relation (\ref{eq:5}) is
replaced with the discrete one\footnote{For  $n > {\cal E}-1/2$ one has
$P_n^{(\mathrm{d})} = \frac{2i}{\Delta}\mathrm{arcsh}
  \frac{|P_n|\Delta}{2\sqrt{1 - |P_n|^2\Delta^2/12}}$ }, 
$$
P_n \to P_n^{(\mathrm{d})} = \frac{2}{\Delta}\mathrm{arcsin}
\frac{P_n\Delta}{2\sqrt{1 + P_n^2\Delta^2/12}}\;.
$$ 
Consequently, one rewrites Eqs. (\ref{eq:2}), (\ref{eq:3}) as
\begin{align}
\notag
&\psi_{n,N_x+1} - \mathrm{e}^{i P_n^{(\mathrm{d})}\Delta}\psi_{n,N_x} =
\delta_{n,{\cal N}} \mathrm{e}^{-iP_n^{(\mathrm{d})}L}(1 - \mathrm{e}^{2i
  P_n^{(\mathrm{d})}\Delta})\;,\\ 
&\label{eq:12}
  \psi_{n,0} - \mathrm{e}^{-i P_n^{(\mathrm{d})}\Delta} 
\psi_{n,-1} = 0\;.
\end{align}
These relations together with Eq. (\ref{eq:10}) form a system of
$N_y(N_x+3)$ linear equations for the same number of unknowns
$\psi_{n,k}$. After solving them,  one calculates the
reflection probability by making use of the discrete current,
$$
{\cal P} = \frac{|J^{(\mathrm{d, out})}|}{|J^{(\mathrm{d,in})}|} \;,
$$
where $J^{(\mathrm{d,in})}$ and $J^{(\mathrm{d,out})}$ are the 
incoming and outgoing 
  currents at $k=N_x+1$,
\begin{align*}
&J^{(\mathrm{d,out})}= \frac{1}{\Delta}\sum_{n< {\cal E}-1/2} 
\frac{\sin(P_n^{(\mathrm{d})} \Delta) |r_n^+|^2}
{\left(1 - \frac13\sin^2\frac{P_n^{(d)} \Delta}{2}\right)^2}\;,\\
&J^{(\mathrm{d,in})} = -\frac{1}{\Delta}
\frac{\sin(P_{\cal N}^{(\mathrm{d})} \Delta)}
{\left(1 - \frac13\sin^2\frac{P_{\cal N}^{(d)} \Delta}{2}\right)^2}\;,
\end{align*}
 while the reflection amplitudes $r_n^+$ are extracted from the
 wave function,
$$
r_n^+ = \frac{\psi_{n,N_x+1} - \mathrm{e}^{-i P_n^{(\mathrm{d})} \Delta}
\psi_{n,N_x}}{1 - \mathrm{e}^{-2iP_n^{(\mathrm{d})} \Delta}}
\cdot\mathrm{e}^{-iP_n^{(\mathrm{d})}L}\;.
$$

We will see shortly that the finite difference approximation works well 
only if the numbers of the lattice points and oscillator levels are
large. Typically, $N_x\sim 10000$, $N_y \sim 500$, and the system
(\ref{eq:12}), (\ref{eq:10}) contains $N_y (N_x+3) \sim 5\cdot 10^{6}$
equations. Such an enormous system of equations cannot be solved with the
general algorithms of linear algebra. So, we took advantage of the
special form of Eqs. (\ref{eq:10}). Namely, the $k$-th matrix equation
relates the vector $\psi_k$ to the unknowns at the adjacent
sites $\psi_{k-1}$ and $\psi_{k+1}$ only; by performing numerically the
matrix inversion it can be recast in the form 
$$
\psi_k = L_k \psi_{k-1} + R_k \psi_{k+1}\;,
$$
where $L_k$ an $R_k$ are the $N_y\times N_y$ matrices. One
substitutes the above formula into the other equations of the system
(\ref{eq:10}), thus excluding
$\psi_k$, as well as the $k$--th matrix
equation. Performing this operation repeatedly, one ends
up with a few matrix equations, which can be solved in a
straightforward manner
by the $LU$ decomposition method. It is worth pointing out that the
variables $\psi_k$ and $\psi_q$ which are not neighbors to each other,
can be excluded in parallel, so that the above algorithm is suitable for 
the multiprocessor machines or computational clusters. The
reader interested in the details of the algorithm should address 
Refs.~\cite{Bonini:1999kj}, \cite{Levkov:2002qg}, or our Fortran
90 code \cite{program}, which hopefully can be executed on other machines. 
 
Before proceeding to the actual calculations, one makes sure that the
parameters of the lattice are chosen properly, 
so that the truncation and discretization errors
are kept under control. Our purpose 
is to get the quantum mechanical results in the semiclassical 
region $g^2\to 0$. Therefore, it is convenient to account explicitly for
the dependence of the lattice parameters on $g$.
The truncation $L$ of the translatory coordinate is fixed by
the condition that the interaction represented by the function $\tilde
a(X)$ is 
small enough at $X=L$. Taking into account the scaling (\ref{newa}) of
$\tilde a(X)$ with $g$, one
obtains the formula  
$$
L = \tilde{L}/g\;,
$$
where $\tilde{L}$ is fixed and large. In the practical calculations we
used the value $\tilde{L} = 12$, which is large enough as
$a(\tilde{L}) \sim 10^{-31}$: at $g>0.07$ this number is 
smaller than the absolute values of the reflection amplitudes,
the latter exceeding $10^{-8}$. We have chosen the truncation
of the oscillator levels in accordance with the condition that the
occupation number of the last mode is negligible,
$$
\sum_k |\psi_{N_y-1,k}| < 10^{-30}\;;
$$
this inequality was satisfied with $N_y$ typically
ranging in between $200$ and $500$. 
The last parameter, the lattice
spacing $\Delta$, should be several times smaller than the minimal De
Broglie wavelength; we have found that the formula 
$$
\Delta = 0.3 \cdot \min_n \frac{1}{|P_n|}
$$
works well enough producing relative errors of order
$10^{-4}$. 

The fact that the discretization 
corresponds to the {\it relative} rather than absolute errors can be
understood as follows. The equations (\ref{eq:10}) and
(\ref{eq:12}) may be 
regarded as the ones describing reflection of a quantum particle
in a kind of crystal. Indeed, the substitution
  $\phi_k=\left(1-\frac{\Delta^2}{12}A_k\right)\psi_k$ brings
  Eq.~(\ref{eq:10}) into the form ${\cal H}^{(d)}\phi=0$, where $\phi$
  is the column composed of $\phi_k$, and ${\cal H}^{(d)}$ is a
  Hermitean linear operator. 
The probability of reflection in a crystal is
exponentially small due to the same dynamical reasons as in the
continuum case. One concludes that the finite value of $\Delta$ gives
rise to corrections
to the suppression exponent, rather than to the reflection
probability itself, i.e. it produces relative discretization
errors. 
We have checked the above physical considerations by performing 
calculations on the lattices with different cutoffs and lattice
spacings. The overall conclusion is that, indeed, the discretization
effects result in relative errors of order $10^{-4}$, while the truncation
errors are always negligible. We kept the round--off errors
under control by exploiting the current conservation law
(\ref{eq:11}), which was checked to hold with precision better than
$10^{-12}$.


\begin{thebibliography}{99}

\bibitem{Perelomov}
A.~M.~Perelomov, V.~S.~Popov and M.~V.~Terent'ev, ZHETF {\bf 51}, 309
(1966).\\
V.~S.~Popov, V.~Kuznetsov and A.~M.~Perelomov, ZHETF {\bf 53}, 331
(1967). 

\bibitem{Miller}
W.~H. Miller,
\newblock Adv. Chem. Phys. {\bf 25}, 69 (1974).

\bibitem{Coleman:1977py}
  I.~Y.~Kobzarev, L.~B.~Okun and M.~B.~Voloshin,
  Sov.\ J.\ Nucl.\ Phys.\  {\bf 20}, 644 (1975)
  [Yad.\ Fiz.\  {\bf 20}, 1229 (1974)].\\
  S.~R.~Coleman,
  Phys.\ Rev.\ D {\bf 15}, 2929 (1977)
  [Erratum-ibid.\ D {\bf 16}, 1248 (1977)].

\bibitem{Heller:1981}
  M.~Davis and E.~Heller, J.Chem.Phys. {\bf 75}, 246 (1981).

\bibitem{Wilkinson:Takada}
  M.~Wilkinson, Physica {\bf 21D}, 341 (1986). \\
  S.~Takada and H.~Nakamura, J. Chem. Phys. {\bf 100}, 98 (1994). \\
  S.~Takada, P.N.~Walker and M.~Wilkinson, Phys. Rev. A {\bf 52}, 
  3546 (1995). \\
  S.~Takada, J. Chem. Phys. {\bf 104}, 3742 (1996).

\bibitem{Banks:1973ps}
  T.~Banks, C.~M.~Bender and T.~T.~Wu,
  Phys.\ Rev.\ D {\bf 8}, 3346 (1973).\\
  T.~Banks and C.~M.~Bender,
  Phys.\ Rev.\ D {\bf 8}, 3366 (1973).

\bibitem{Rubakov:1992ec}
  V.~A.~Rubakov, D.~T.~Son and P.~G.~Tinyakov,
  Phys.\ Lett.\ B {\bf 287}, 342 (1992).

\bibitem{Kuznetsov:1997az}
  A.~N.~Kuznetsov and P.~G.~Tinyakov,
  Phys.\ Rev.\ D {\bf 56}, 1156 (1997)
  [arXiv:hep-ph/9703256].

\bibitem{Bezrukov:2003er}
  F.~Bezrukov, D.~Levkov, C.~Rebbi, V.~A.~Rubakov and P.~Tinyakov,
  Phys.\ Rev.\ D {\bf 68}, 036005 (2003)
  [arXiv:hep-ph/0304180]; Phys.\ Lett.\ B {\bf 574}, 75 (2003)
  [arXiv:hep-ph/0305300].

\bibitem{Levkov:2004tf}
  D.~G.~Levkov and S.~M.~Sibiryakov,
  Phys.\ Rev.\ D {\bf 71}, 025001 (2005)
  [arXiv:hep-th/0410198];
  JETP Lett.\  {\bf 81}, 53 (2005)
  [Pisma Zh.\ Eksp.\ Teor.\ Fiz.\  {\bf 81}, 60 (2005)]
  [arXiv:hep-th/0412253].

\bibitem{Affleck:1980mp}
  I.~Affleck, Nucl.\ Phys.\ B {\bf 191}, 429 (1981).

\bibitem{Berry:1972}
M.~V.~Berry and K.~E.~Mount, Rep.\ Prog.\ Phys. {\bf 35}, 315 (1972). 

\bibitem{Adachi:1986}
S.~Adachi, Ann.Phys. {\bf 195}, 45 (1989).

\bibitem{Shudo:1996}
A.~Shudo, K.~S.~Ikeda, Phys.Rev.Lett. {\bf 76}, 4151 (1996).

\bibitem{Ribeiro:2004}
A.D.~Ribeiro, M.A.M.~de~Aguiar, M.~Baranger, Phys.\ Rev.\ E {\bf 69},
066204 (2004).
 
\bibitem{Parisio:2005}
F.~Parisio, M.A.M.~de~Aguiar, J.\ Phys.\ A: Math.\ Gen. {\bf 38},
9317 (2005). 

\bibitem{Bohigas1993}
O.~Bohigas, S.~Tomsovic and D.~Ullmo,
Phys.\ Rep.\ {\bf 223}, 43 (1993).

\bibitem{Doron1995}
E.~Doron and S.~D.~Frischat,
Phys.\ Rev.\ Lett. {\bf 75}, 3661 (1995);\\
S.~D.~Frischat and E.~Doron,
Phys.\ Rev.\ E {\bf 57}, 1421 (1998).

\bibitem{Mouchet2001}
A.~Mouchet, C.~Miniatura, R.~Kaiser, B.~Gr\'{e}maud, D.~Delande,
Phys.\ Rev.\ E {\bf 64}, 016221 (2001).

\bibitem{Shudo1995_1} 
  A.~Shudo, K.S. Ikeda, Phys.\ Rev.\ Lett.\ {\bf 74}, 682 (1995);
  Physica { \bf D 115}, 234 (1998).

\bibitem{Shudo2002}
A.~Shudo, Y.~Ishii and K.~S.~Ikeda,
J.\ Phys.\ A {\bf 35}, L225 (2002).

\bibitem{Onishi2003}
T.~Onishi, A.~Shudo, K.S.~Ikeda and K.~Takahashi, Phys.\ Rev.\ E 
{\bf 68}, 056211 (2003). 

\bibitem{Heller2002}
  T.Van Voorhis, E.J.~Heller, Phys.\ Rev.\ A {\bf 66}, 050501(R)
  (2002). 

\bibitem{Eckhardt1986}
B.~Eckhardt, C.~Jung J.\ Phys.\ A {\bf 19}, L829 (1986);\\
B.~Eckhardt, Physica\ D {\bf 33}, 89 (1988);\\
P.~Gaspard, S.A.~Rice, J.\ Chem.\ Phys. {\bf 90}, 2225 (1988). 

\bibitem{Zakharov:1990xt}
  V.~I.~Zakharov, Nucl.\ Phys.\ B {\bf 353}, 683 (1991); 
  Phys.\ Rev.\ Lett.\  {\bf 67}, 3650 (1991).\\
  G.~Veneziano,
  Mod.\ Phys.\ Lett.\ A {\bf 7}, 1661 (1992).

\bibitem{Maggiore:1991kh}
  M.~Maggiore and M.~A.~Shifman,
  Phys.\ Rev.\ D {\bf 46}, 3550 (1992).

\bibitem{Voloshin:1993dk}
  M.~B.~Voloshin,
  Phys.\ Rev.\ D {\bf 49}, 2014 (1994).

\bibitem{Rubakov:1994hz}
  V.~A.~Rubakov and D.~T.~Son,
  Nucl.\ Phys.\ B {\bf 424}, 55 (1994)
  [arXiv:hep-ph/9401257].

\bibitem{Drew2005}
C.~S.~Drew, S.~C.~Creagh and R.~H.~Tew,
Phys.\ Rev.\ A {\bf 72}, 062501 (2005).

\bibitem{Bezrukov:2003yf}
  F.~Bezrukov and D.~Levkov,  arXiv:quant-ph/0301022; 
  J.\ Exp.\ Theor.\ Phys.\  {\bf 98}, 820 (2004)
  [Zh.\ Eksp.\ Teor.\ Fiz.\  {\bf 125}, 938 (2004)]
  [arXiv:quant-ph/0312144].

\bibitem{Takahashi2003}
K.~Takahashi and K.S.~Ikeda, J.\ Phys.\ A {\bf 36}, 7953 (2003);
Europhys. \ Lett. \ {\bf 71}, 193 (2005).

\bibitem{Klinkhamer:1984di}
  F.~R.~Klinkhamer and N.~S.~Manton,
  Phys.\ Rev.\ D {\bf 30}, 2212 (1984).


\bibitem{Bonini:1999kj}
  G.~F.~Bonini, A.~G.~Cohen, C.~Rebbi and V.~A.~Rubakov,
  Phys.\ Rev.\ D {\bf 60}, 076004 (1999)
  [arXiv:hep-ph/9901226];  arXiv:quant-ph/9901062.

\bibitem{prefactor}
D.~G.~Levkov, A.~G.~Panin and S.~M.~Sibiryakov,
  ``Tunneling via unstable semiclassical solutions,''
  arXiv:0707.0433 [quant-ph].

\bibitem{wigles}
D.~G.~Levkov, A.~G.~Panin and S.~M.~Sibiryakov,
Phys.\ Rev.\ A {\bf 76}, 032114 (2007) [arXiv:0704.0409].

\bibitem{NumericalRecipes}
W.H.~Press, S.A.~Teukolsky, W.T.~Vetterling, B.P.~Flannery,
``Numerical recipes in C: the art of scientific computing'', {\it
  Cambridge University Press}, 1992.

\bibitem{Xavier1996}
A.L.~Xavier Jr., M.A.M.~de~Aguiar, Ann.\ Phys {\bf 252}, 458 (1996).

\bibitem{Levkov:2002qg}
  D.~Levkov, C.~Rebbi and V.~A.~Rubakov, Phys.\ Rev.\ D {\bf 66},
  083516 (2002) [arXiv:gr-qc/0206028].

\bibitem{program} {\it http://solver.inr.ac.ru}

\end{thebibliography}
\end{document}